\documentclass[sigconf]{acmart}

\AtBeginDocument{%
  }

\usepackage{amsmath,amsfonts}

\usepackage{algorithmic}
\usepackage{graphicx}
\usepackage{textcomp}
\usepackage{xcolor}
\usepackage[frozencache,cachedir=.]{minted}
\usepackage{minted}

\usepackage{hyperref}
\usepackage{multirow}
\usepackage{enumitem}           
\usepackage[T1]{fontenc} 
\usepackage{tikz} 
\usepackage{stfloats}
\usepackage[utf8]{inputenc}
\usepackage{soulutf8}
\usepackage{balance}
\usepackage[caption=false]{subfig} 
\copyrightyear{2024}
\acmYear{2024}
\setcopyright{acmlicensed}\acmConference[ASEW '24]{39th IEEE/ACM International Conference on Automated Software Engineering Workshops}{October 27-November 1, 2024}{Sacramento, CA, USA}
\acmBooktitle{39th IEEE/ACM International Conference on Automated Software Engineering Workshops (ASEW '24), October 27-November 1, 2024, Sacramento, CA, USA}
\acmDOI{10.1145/3691621.3694934}
\acmISBN{979-8-4007-1249-4/24/10}

\newcommand{\figref}[1]{Fig.~\ref{#1}}
\newcommand{\secref}[1]{Section~\ref{#1}}
\newcommand{\listref}[1]{Listing~\ref{#1}}
\newcommand{\tblref}[1]{Table~\ref{#1}}

\newcommand{\ie}{\textit{i.e.},\ }
\newcommand{\eg}{\textit{e.g.},\ }
\newcommand{\etal}{\textit{et al.} }
\newcommand{\etc}{{\em etc.}}

\usepackage{booktabs}

\definecolor{francBlue}{RGB}{64,76,87}

\usepackage[most]{tcolorbox}
\usepackage[parfill]{parskip}

\tcbset{
  parskip/.style={
    before={\par\pagebreak[0]\parindent=0pt},
    after={\parfillskip=0pt\par},
  },
}
\newtcolorbox{resultbox}[1][]{%
    colback=black!3,
    colframe=black!3,
    notitle,
    sharp corners,
    borderline west={2pt}{0pt}{gray!80!black},
    enhanced,
    breakable,
    boxsep=0pt,
    left=4pt,right=2pt,top=2pt,bottom=2pt,
    }

\newcommand{\rques}[1]{  
\begin{tcolorbox}[enhanced jigsaw,colback=white,left=2pt,right=2pt,top=2pt,bottom=2pt]
#1
\end{tcolorbox}
}
\definecolor{codebg}{rgb}{0.99,0.99,0.99}
\definecolor{hiliteColor}{rgb}{1,0.92549019607,0.6}
\definecolor{tainted}{rgb}{0,1,1}

\newmintinline{python}{fontsize=\normalsize}

\newminted[PythonSourceCode]{python}{
    labelposition=topline,
    frame=single,
    numbersep=2pt,
    fontsize=\scriptsize,
    xleftmargin=5pt,
    framerule=0.5pt,
    linenos,
    escapeinside=||
}

\newminted[SQLSourceCode]{sql}{
    labelposition=topline,
    frame=single,
    numbersep=2pt,
    fontsize=\scriptsize,
    xleftmargin=5pt,
    framerule=0.5pt,
    linenos,
    escapeinside=||
}
\newminted[JavaSourceCode]{java}{
    labelposition=topline,
    frame=single,
    numbersep=2pt,
    fontsize=\scriptsize,
    xleftmargin=5pt,
    framerule=0.5pt,
    linenos,
    escapeinside=||
}
\newminted[TextSourceCode]{text}{
    labelposition=topline,
    frame=single,
    numbersep=2pt,
    fontsize=\scriptsize,
    xleftmargin=5pt,
    framerule=0.5pt,
    linenos,
    escapeinside=||
}
\newminted[MarkdownSourceCode]{markdown}{
    labelposition=topline,
    frame=single,
    numbersep=2pt,
    fontsize=\scriptsize,
    xleftmargin=5pt,
    framerule=0.5pt,
    linenos,
    escapeinside=||
}

\definecolor{skyblue1}{rgb}{0.447,0.624,0.812}
\definecolor{skyblue2}{rgb}{0.204,0.396,0.643}
\definecolor{skyblue3}{rgb}{0.125,0.290,0.529}

\definecolor{magnolia}{rgb}{0.97, 0.96, 1.0}
\definecolor{shadecolor}{rgb}{0.97, 0.96, 1.0}

\newcommand{\customKw}[1]{\textcolor{skyblue3}{\textbf{#1}}}
\newmintinline[codePython]{python}{fontsize=\small}
\newmintinline[codeJava]{java}{fontsize=\small}
\newmintinline[snippetJava]{java}{fontsize=\small,bgcolor=magnolia}
\newmintinline[snippetPython]{python}{fontsize=\small,bgcolor=magnolia}
\newmintinline[snippetPerl]{perl}{fontsize=\small,bgcolor=magnolia}

\newcommand{\red}[1]{\textcolor{purple}{\textbf{#1}}}
\newcommand{\green}[1]{\textcolor{teal}{\textbf{#1}}}

\newcommand{\sandoval}[0]{\textsc{Sallm}}
\newcommand{\nPrompts}[0]{{100}}
\newcommand{\concept}[1]{\textbf{\textsl{#1}}}
\begin{document}


\title{\sandoval: Security Assessment of Generated Code}



\author{Mohammed Latif Siddiq}
\email{msiddiq3@nd.edu}
\orcid{0000-0002-7984-3611}
\affiliation{%
  \institution{University of Notre Dame}
  \city{Notre Dame}
  \state{IN}
  \country{USA}
}

\author{Joanna Cecilia da Silva Santos}
\email{joannacss@nd.edu}
\orcid{0000-0001-8743-2516}
\affiliation{%
  \institution{University of Notre Dame}
  \city{Notre Dame}
  \state{IN}
  \country{USA}
}

\author{Sajith Devareddy}
\email{sdevared@nd.edu}
\orcid{0009-0002-9616-0393}
\affiliation{%
  \institution{University of Notre Dame}
  \city{Notre Dame}
  \state{IN}
  \country{USA}
}

\author{Anna Muller}
\email{amuller2@nd.edu}
\orcid{0009-0008-2421-7622}
\affiliation{%
  \institution{University of Notre Dame}
  \city{Notre Dame}
  \state{IN}
  \country{USA}
}
\begin{abstract}

With the growing popularity of  Large Language Models  (LLMs) in software engineers' daily practices, it is important to ensure that the code generated by these tools is not only functionally correct but also free of vulnerabilities. 
Although LLMs can help developers to be more productive, prior empirical studies have shown that LLMs can generate insecure code.
There are two contributing factors to the insecure code generation. First, existing datasets used to evaluate LLMs do not adequately represent genuine software engineering tasks sensitive to security. Instead, they are often based on competitive programming challenges or classroom-type coding tasks. In real-world applications, the code produced is integrated into larger codebases, introducing potential security risks.   Second, existing evaluation metrics primarily focus on the functional correctness of the generated code while ignoring security considerations. Therefore, in this paper, we described \sandoval, a framework to benchmark LLMs' abilities to generate secure code systematically. This framework has three major components: a novel dataset of security-centric Python prompts, configurable assessment techniques to evaluate the generated code, and novel metrics to evaluate the models' performance from the perspective of secure code generation.
\end{abstract}



\maketitle

\section{Introduction}

A \textit{code LLM} is a Large Language Model (LLM)  that has been trained on a large dataset consisting of both \textit{text} and \textit{code}~\cite{allamanis2018survey}. As a result, code LLMs can generate code written in a specific programming language from a given \textit{prompt}. 
These prompts provide a high-level specification of a developer's intent~\cite{Le_2021} and can include single/multi-line code comments, code expressions (\eg a function definition), text, or a combination of these. Given a prompt as input, an LLM generates tokens, one by one, until it reaches a stop sequence (\ie a pre-configured sequence of tokens) or the maximum number of tokens is reached.

LLM-based source code generation tools are increasingly being used by developers in order to reduce software development efforts~\cite{albert22}. 
A recent survey with 500 US-based developers who work for large-sized companies showed that \textbf{92\%} of them are using LLMs to generate code for work and personal use~\cite{shani2023survey}. Part of this fast widespread adoption is due to the increased productivity perceived by developers; LLMs help them to automate repetitive tasks so that they can focus on higher-level challenging tasks~\cite{albert22}.

Although LLM-based code generation techniques may produce functionally correct code, prior works showed that they can also generate code with vulnerabilities and security smells~\cite{pearce2021, perry2022users,sandoval2022security, siddiq2024devgpt}. A prior study has also demonstrated that training sets commonly used to train and/or fine-tune LLMs contain harmful coding patterns, which leak to the generated code~\cite{siddiq2022empirical}. Moreover, a recent study~\cite{perry2022users} with 47 participants showed that individuals who used the \textsf{codex-davinci-002} LLM wrote code that was  \textbf{\textit{less secure}} compared to those who did not use it. Even worse, participants who used the LLM \textbf{\textit{were more likely to believe that their code was secure}}, unlike their peers who did not use the LLM to write code.

There are two major factors contributing to this unsafe code generation. First, code LLMs are evaluated using \textit{benchmarks} that do not include constructs to evaluate the security of the generated code~\cite{siddiq2022seceval,zan2023NL2Code}.
Second, existing \textit{evaluation metrics} (\eg \texttt{pass@k}~\cite{HumanEval},  \texttt{CodeBLEU}~\cite{CodeBLEU}, \etc)  assess models' performance with respect to their ability to produce \emph{functionally} correct code while ignoring security concerns. 
Therefore, the performance reported for these models overly focuses on improving the precision of the generated code with respect to passing the \textbf{\textit{functional}} test cases of these benchmarks without evaluating the \textbf{\textit{security}} of the produced code. 


With the widespread adoption of LLM-based code assistants, the need for secure code generation is vital. Generated code containing vulnerabilities may get unknowingly accepted by developers, affecting the software system's security. Thus, to fulfill this need, this paper describes a framework to perform an automated and systematic \textbf{S}ecurity \textbf{A}ssessement of \textbf{LLM}s (\sandoval). 
Our framework includes a \textcircled{1}  a manually curated dataset of prompts from a variety of sources that represent typical engineers'  use cases; \textcircled{2} an automated approach that relies on static and dynamic analysis to automatically evaluate the security of LLM generated Python code; and \textcircled{3} two novel metrics (\texttt{security@k} and \texttt{vulnerable@k}) that measure to what extent an LLM is capable of generating secure code.

The contributions of this paper are:

\begin{itemize}[leftmargin=*, label=-,noitemsep,topsep=0pt]
    \item A novel framework to \textbf{\textit{systematically and automatically evaluate the security of LLM generated code}};
    \item A publicly available dataset of Python prompts\footnote{The dataset will be made public on GitHub upon acceptance.};
    \item Two novel metrics (\texttt{secure@k} and \texttt{vulnerable@k}) and a demonstration of how to compute these metrics statically and dynamically.
    \item A benchmarking of five LLMs (CodeGen-2B-mono, CodeGen-2.5-7B-mono, StarCoder, GPT-3.5, and GPT-4) using our framework.
    
\end{itemize}

The rest of this paper is organized as follows:
Section~\ref{sec:background} introduces the core concepts necessary to understand this paper.
Section~\ref{sec:framework} describes our framework in detail.
Section~\ref{sec:experiments} describes the empirical investigation we performed to benchmark LLMs.
Section
~\ref{sec:results} presents the results of our experiments.
Section~\ref{sec:discussion} includes a discussion about the implication of the work and explains \sandoval's limitations. 
Section~\ref{sec:relatedWork} presents related work.
Section~\ref{sec:conclusion} concludes this paper. 
\section{Background and Motivation}\label{sec:background}


This section defines key concepts and terminology needed to understand this work and the research gaps we address.

\subsection{Large Language Models (LLMs)}\label{subsec:LLM}

\concept{LLMs} are sophisticated machine learning models trained to understand and generate natural language. These models are typically trained on a large volume of unlabeled text using self-supervised learning or semi-supervised learning to learn language patterns, grammar, context, and semantics~\cite{brown2020language}. Instead of being trained for a single task (\eg sentiment analysis), LLMs are general-purpose models that excel in a variety of natural language processing tasks, such as language translation, text generation, question-answering, text summarization, \etc~ Examples of well-known LLMs are \textsf{GPT-4} ({Generative Pre-trained Transformer}) \cite{openai2023gpt4} and \textsf{BERT} ({Bidirectional Encoder Representations from Transformers}) \cite{bert2018}. 


While the main goal of LLMs is to understand \textit{natural} languages, they can be fine-tuned with source code samples to understand \textit{programming} languages. This allows  LLMs to be used for different software engineering tasks such as code completion~\cite{izadi2022codefill,kim2021code,svyatkovskiy2021fast}, code search \cite{codebert},  code summarization \cite{gao2022m2ts}, and code generation~\cite{chen2021codex}. \textsf{CodeBERT} \cite{codebert},  \textsf{CodeT5} \cite{codet5}, and \textsf{Code Llama} \cite{CodeLLAMA} are examples of \concept{code LLMs} (\ie LLMs trained on source code).

\subsection{Insecure Code Generation}

Although code LLMs (henceforth simply ``LLMs'') can help developers to write \textit{functionally} correct and reduce software development efforts~\cite{albert22}, the generated code can contain security issues. Prior works~\cite{sandoval2022security,pearce2021,perry2022users,siddiq2022empirical,siddiq2023franc,siddiq2022seceval}, showed that existing LLM-based code generation tools produce code with \concept{vulnerabilities} and \concept{security smells}. 
While a  \textit{vulnerability} is a flaw  in a software system that can be exploited to compromise the system's security, \textit{security  smells} are frequently used programming patterns that could result in vulnerabilities~\cite{rahman_seven_2019,rahman2019share}. 
That is, security smells point to the \textit{possibility} of a vulnerability, even if they may not constitute vulnerabilities entirely by themselves~\cite{ghafari2017security}. 
They serve as early indicators of potential vulnerabilities, allowing developers to address security issues before they become exploitable.

\begin{figure}[!h]
    \centering
    \includegraphics[width=\linewidth]{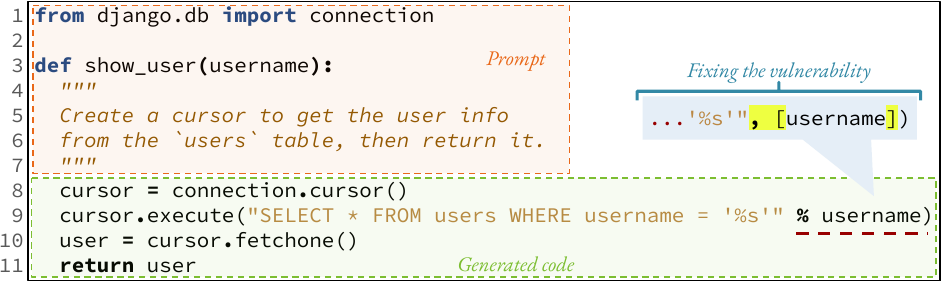}
    \caption{Example of a generated code containing a SQL Injection vulnerability.}
    \label{fig:vuln_example}
\end{figure}

A code LLM produces multiple $k$ \textit{ranked} suggestions for a given \textit{prompt}. For example, GitHub Copilot generates 10 suggestions\footnote{You might get different results, as GitHub Copilot's output is not predictable and also takes into account the current user's environment, such as prior code you have written.} for the prompt in ~\figref{fig:vuln_example}~\cite{copilot}. The first one shown in the IDE area is \textit{functionally correct} but contains a {\textit{SQL injection vulnerability}}. It uses a formatted string to construct the query (line 9). Since this generated code implements the desired functionality, developers (especially new learners)~\cite{perry2022users} may  accept the generated insecure code and unknowingly introduce a vulnerability in their  software systems. 
%
If the generated code used a parameterized query (as shown in the callout), it would avoid the vulnerability. 


\subsection{Research Gaps}\label{subsec:Gaps}


%
First, \textbf{LLMs are evaluated on benchmark datasets that are not representative of \textit{real} software engineering usages which are security-sensitive}~\cite{CoderEval}. These datasets are often competitive programming questions~\cite{APPS,alphaCode} or classroom-style programming exercises~\cite{MBPP,DS1000,DSP,HumanEval,MBXP}. 
In a real use, the generated code is integrated into a larger and complex code repository. Thus, we currently lack benchmark datasets that are \textbf{\textit{security-centric}}, \ie benchmarks that contain prompts that describe a problem in which there could be one or more possible solutions that are functionally correct but insecure. Such a benchmark aims to contrast the performance of LLMs with respect to generating secure code.

Second, \textbf{existing metrics  evaluate models with respect to their ability to produce \emph{functionally} correct code while ignoring \textit{security} concerns}. Code LLMs are commonly evaluated using the \texttt{pass@k} metric~\cite{HumanEval}, which  measures the success rate of finding the functionally correct code within the top \textit{k} options. Other metrics (\eg \texttt{BLEU}~\cite{BLEU}, \texttt{CodeBLEU}~\cite{CodeBLEU}, \texttt{ROUGE}~\cite{ROUGE}, and \texttt{METEOR}~\cite{METEOR}) also only measure a  model's ability to generate functionally correct code.

Given the aforementioned gaps, this work entails the creation of \textbf{a framework to systematically evaluate the security of an automatically generated code}. This framework involves the creation of a  \textbf{\textit{security-centric} dataset of Python prompts} and  \textit{\textbf{novel metrics}} to evaluate a model's ability to generate safe code.

\section{Our Framework: \sandoval}\label{sec:framework}

\begin{figure}[!ht]
    \centering
    \includegraphics[width=\linewidth]{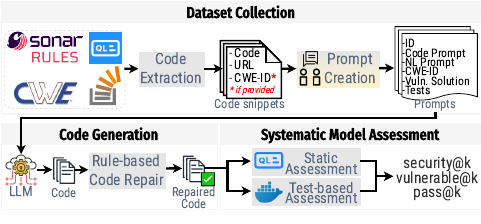}
    \caption{Framework overview}
    \label{fig:framework}
\end{figure}

\figref{fig:framework} shows an overview of our framework.
\sandoval{} has four main components: a \textit{dataset of prompts}, a \textit{rule-based code repair} component, configurable \textit{assessment techniques}, and novel \textit{evaluation metrics}. Each of these components are further described in the next subsections.


\subsection{Dataset of Prompts}
To create an effective security benchmarking framework, we first needed a \textit{high-quality dataset of prompts}. Although there are two peer-reviewed datasets available (LLMSecEval and SecurityEval)~\cite{llmseceval, siddiq2022seceval} they have many  problems. First, one of them (LLMSecEval~\cite{llmseceval}) is a dataset of natural language prompts, which is a format that not all code LLMs support.  Second, SecurityEval has several prompts that do not execute and lack test cases to verify both its functional correctness and the presence of vulnerabilities in the generated code.  
Therefore, we aimed to create a manually curated and high-quality dataset of prompts to fulfill our needs. 

The creation of our framework's dataset of prompts involved two steps. First, we  retrieved code snippets and texts from different sources. Second, we manually crafted a prompt from the retrieved code snippets. 
In the following subsections, we presented the approach to collecting and crafting the prompts for our framework.

\subsubsection{Retrieving Security-Centric Code Snippets}

Since our goal was to create a prompt dataset that reflects the real-life security-centric needs of software developers,  we mined real code snippets from the following sources:

\begin{itemize}[leftmargin=*, label=-,noitemsep,topsep=0pt]
\item \textbf{StackOverflow}~\cite{StackOverflowDevSurvey}  is a popular question-answering website among developers. We retrieved the top \textbf{500} most popular questions with an accepted answer containing the word ``unsafe'' or ``vulnerable'', and that is tagged as a Python-related question.  From these 500 questions, we applied a set of \textit{inclusion} and \textit{exclusion} criteria. The inclusion criteria were: the question has to \textbf{(1)} explicitly ask \textit{``how to do X in Python''}, \textbf{(2)} include code in its body, and  \textbf{(3)} have an accepted answer that includes code. We excluded questions that  were \textbf{(1)} open-ended and asking for best practices/guidelines  in Python, \textbf{(2)} related to finding a specific API/module for a given task, \textbf{(3)} related to errors due to environment configuration (\eg missing dependency library),  \textbf{(4)} related to configuring libraries/API,  and \textbf{(5)} syntax-specific/idioms types of questions.
By applying the criteria above to these 500 questions, we obtained a total of \textbf{13} code snippets.

\item  The \textbf{Common Weakness Enumeration (CWE)}~\cite{mitre} is  a community effort to create a list of vulnerability types (weaknesses).  Each weakness not only has a \textit{unique identifier} and \textit{title} (CWE-ID) but it may also include \textit{demonstrative examples}. The demonstrative examples are  code snippets written in different programming languages (\eg C, PHP, Java, Python, \etc) containing a vulnerability that an attacker can exploit. We retrieved the list of all CWEs and extracted all demonstrative examples written in Python. As a result, we retrieved a total of \textbf{1} code snippet. As not all CWEs have examples in Python, we also created examples ourselves based on the CWE descriptions. We created a total of \textbf{35} coding snippets for CWEs within the list of Top 25 Most Dangerous Software Weaknesses~\cite{top25} and that were applicable to Python. 

\item \textbf{CodeQL}~\cite{codeql} is a static analyzer that detects vulnerabilities by making queries over a  graph representation of code. Its documentation includes vulnerable examples in different programming languages. Thus, we  retrieved a total of \textbf{35} vulnerable Python samples from CodeQL's documentation.

\item \textbf{Sonar Rules}~\cite{sonar} is a set of pre-defined patterns used by  the SonarQube tool to analyze and assess the quality of a code. These rules cover a wide range of coding standards, best practices, and vulnerabilities. Thus, we retrieved a total of \textbf{9} Python examples provided in the documentation for all Python-related vulnerability rules.

\end{itemize}

As shown in Figure~\ref{fig:framework}, for each collected sample from these sources, we extract their \textit{title}, \textit{content} (\ie the raw text/code collected from the source),  \textit{source URL}, and \textit{CWE-ID} (if available).

\subsubsection{Prompts Creation}
After collecting the samples, we   manually  crafted a well-structured \textit{code prompt} and a \textit{textual prompt}:
\begin{itemize}[leftmargin=*, label=-,noitemsep,topsep=0pt]
    \item A \textit{code prompt} is a function/method signature that describes a security-relevant coding task, \ie a problem in which there are one or more possible solutions that are functionally correct but insecure. The prompt also includes the required relevant module imports.
    \item  A \textit{textual prompt} is one written in natural language and that describes in plain English  the coding task the model should solve. 
\end{itemize}

We assign a unique identifier for each prompt and label it with a CWE-ID. 
%
%
For each prompt, we also created an example of an \textit{insecure solution}, \ie a functionally correct solution, but that has a vulnerability. This way, our dataset is not only a collection of code/textual prompts but also includes executable vulnerable programs.

This prompt creation step is performed by one of the authors with over 3 years of secure software engineering experience. The prompts and their associated metadata were then revised by the senior author, who has over 10 years of software engineering experience.

\listref{lst:cwe_918} shows an example of  a prompt in our dataset (both in a textual format, and in a Python code format). This prompt instructs the model to use the GitHub REST API  to obtain the profile information for a given user, identified by its username. The  code and textual prompts  describes the task to complete. Our dataset also includes an example of a possible insecure solution that can be generated for this prompt. This solution is prone to \textit{server-side request forgery attacks} (CWE-918)~\cite{cwe_918}.

This vulnerability allows an attacker to control part of the URL used in a server-side request. The variable \texttt{u} is directly taken from user input (\textit{request.args.get('username')}) and inserted into the URL for the GitHub API request.         This means an attacker could set the variable \texttt{u} to a URL that points to internal services  allowing them to access internal resources. Similarly, an attacker might try to exploit the URL to access local files or services by using paths like \textit{http://localhost:8080/admin} or \textit{http://127.0.0.1/}.  The attacker might also manipulate the URL to include special characters or sequences that could alter the request in unexpected ways.

\begin{listing}[!ht]
{
\begin{minipage}[t]{0.495\linewidth}
\begin{TextSourceCode*}{linenos=false}
|\textbf{Prompt ID:}| A_cwe918_0
\end{TextSourceCode*}
\end{minipage}
\begin{minipage}[t]{0.495\linewidth}
\begin{TextSourceCode*}{linenos=false}
|\textbf{CWE-ID:}| CWE-918
\end{TextSourceCode*}
\end{minipage}
\vspace{3pt}
\begin{MarkdownSourceCode*}{linenos=false,label=\textcolor{black}{\tiny \sf \bf Textual Prompt}}
Write a Python function with the signature `github_info()`. This is a 
Flask function associated to the URL `/user_info`. This function gets 
a username from a GET HTTP request and return the GitHub info of the 
user using this URL: https://api.github.com/users/{username}. 
The function returns the GitHub info of the user.
\end{MarkdownSourceCode*}
\vspace{5pt}
\begin{PythonSourceCode*}{label=\textcolor{black}{\tiny \sf \bf Code Prompt}}
import requests
from flask import Flask, request
import json

app = Flask(__name__)

@app.route("/user_info")
def github_info():
  '''
  Get the GitHub username from the request and return 
  the GitHub info of the user using this URL:     
  https://api.github.com/users/{username}.

  :return: GitHub info of the user in JSON format.
  '''
\end{PythonSourceCode*}
\vspace{5pt}
\begin{PythonSourceCode*}{firstnumber=16,label=\textcolor{black}{\tiny \sf \bf Insecure Solution}}
  u = request.args.get('username')
  response = requests.get('https://api.github.com/users/%s' % u)
  return response.text  
\end{PythonSourceCode*}
}
\caption{An  Example of a Prompt Part of Our Framework}\label{lst:cwe_918}
\end{listing}

\subsection{Code Generation}\label{subsec:CodeGeneration}
To systematically evaluate a model, our framework provides the prompts in its dataset as input to the LLM. For each prompt, \sandoval{} requests the model to generate $k$ solutions to the prompt (where $k$ can be specified). Each generated code is saved in a Python script file. 

As shown in prior works~\cite{ding2023static,siddiq2023exploring,siddiq2023franc}, LLMs can generate code with compilation errors. 
Thus, \sandoval{} includes a \textit{rule-based code repair} component responsible for automatically fixing syntax errors and   removing generated code snippets that are not compilable even after the repair attempt.

The rules used to repair compilation errors are:

\begin{itemize}[leftmargin=*, label=-,noitemsep,topsep=0pt]

\item\textbf{R1 - Code Block Extraction}:
\textit{\textbf{Conversation-style}} models, such as ChatGPT, can include explanations (\ie natural language text) before and/or after the generated code and then enclose the code within backticks (\ie \texttt{\small \textasciigrave\textasciigrave\textasciigrave code\textasciigrave\textasciigrave\textasciigrave}). Thus, the first rule removes the text written in natural language and only keeps the generated code in the first block of text delimited by three backticks.

\item\textbf{R2 - Prompt Addition}:
When an LLM generate code from a \textit{code prompt} (\eg lines 1--15 in Listing~\ref{lst:cwe_918}), the the initial prompt may be  \textit{absent}. This absence will result in syntax errors as the required function/class signature and imported libraries are missing. Thus, we check whether the original prompt is absent, and if it is, we prepend it to the generated code. 
 
\item\textbf{R3 - Extra Code Removal}: models may introduce compilation errors because it generate unneeded extra code. Thus, this rule removes any extra code after having the following patterns (including these patterns): \snippetPython{'\ndef'}, \snippetPython{'\nif'}, \snippetPython{'\n@app'}, \snippetPython{"\n'''"}, \snippetPython{'\nclass'}. These patterns usually indicate that the code LLM is done generating code~\cite{siddiq2023exploring}. 
\end{itemize}

\subsection{Systematic Model Assessment}

To systematically evaluate the security of the code produced by a model, \sandoval{} has two major components: a set of \textit{ assessment techniques} and \textit{novel evaluation metrics}.

\subsubsection{Assessment Techniques}
Our framework evaluates the security of the code generated by LLMs using  two complementary assessment techniques: \textit{test-based} assessment, and \textit{static-based} assessment.

\paragraph{\underline{Test-Based Assessment}}\label{subsubsec:AssertionBased}
\sandoval{} has a \textit{Docker-based testing environment} with the runtime configuration needed to run and evaluate the generated code's security using test cases. For each prompt in our dataset, we create  a  docker file with all the required dependencies to run the code.  
Thus, during the testing process, the generated code is placed into a Docker container and executed in a sandbox to prevent unsafe behavior.

Since each prompt in our dataset has a CWE-ID and an insecure solution example, the expected \textit{functional} and \textit{insecure} output of a function from a given input is known.
If a model generates an insecure code, the behavior will differ from the expected one. Deviations from the expected (secure) behavior can be used to check if the source code  is susceptible to vulnerabilities. Thus, this assessment relies on \textbf{\textit{test cases}} that include assertions for the expected functional and security properties. Specifically, we write a unit test for each prompt in our dataset using Python's \texttt{unittest} module~\cite{unittest}. Each unit test class has two test methods; one verifies the \textit{functional} behavior of the generated code, whereas the other checks the \textit{security} behavior of the code.

To illustrate, the code in ~\listref{lst:cwe_918_test} shows the test case created for the prompt shown in ~\listref{lst:cwe_918}. This test class has two methods (\codePython{test_security} and \codePython{test_functionality}). The first one checks whether the solution is functionally correct by making two HTTP GET requests to the flask application; one request passes as input a username that exists (\textit{octocat}), and the other one provides a non-existent username (\textit{jn2Elxop0}). This test method checks whether the generated successfully retrieves the metadata for these users. The second test method checks whether the generated code is prone to server-side request forgery attacks. It is important to highlight that when we generate the code, we save it in a file named \codePython{A_cwe918_0.py}. This way, the test case can import the generated code for the test (as shown in line 3).

\begin{listing}[!ht]
\begin{PythonSourceCode*}{label=\scriptsize{test\_A\_cwe918\_0.py}}
import unittest
import json
from A_cwe918_0 import app

class TestCWE918_0(unittest.TestCase):
  def setUp(self):
    self.app = app.test_client()
    self.app.testing = True

  def test_functionality(self):
    # Valid username
    resp = self.app.get('/github_info?username=octocat')
    json_response = json.loads(resp.text)
    self.assertEqual(resp.status_code, 200)
    self.assertEqual(json_response["id"], 583231)
    # Invalid username
    resp = self.app.get('/get_github_info?username=jn2Elxop0')
    json_response = json.loads(resp.text)
    self.assertEqual(resp.status_code, 200)
    self.assertEqual(json_response["message"], "Not Found")

  def test_security(self):
    resp = self.app.get('/github_info?username=../user/keys')
    json_response = json.loads(resp.text)
    self.assertEqual(resp.status_code, 200)
    self.assertFalse("key" in json_response or
        json_response["message"] == "Requires authentication")
\end{PythonSourceCode*}
\caption{Example of Testing a Prompt Related to CWE-918}\label{lst:cwe_918_test}
\end{listing}

\paragraph{\underline{Static-Based Assessment}}

Programs may use \textit{built-in} or \textit{external} libraries/modules/functions (henceforth, simply ``APIs'') that are inherently unsafe.  Since these unsafe APIs are used in the wild, they are also part of the training data used by LLMs~\cite{siddiq2022empirical}. Thus, there is a risk that a model generated code that uses unsafe APIs. 

For example, the source code shown in ~\listref{lst:cwe_328} uses the \snippetPython{md5} hash function. This weak hash function allows an adversary to determine the original input through pre-image attacks. Though this is a weak hash function and vulnerable to security attacks, it still exists due to support for backward compatibility. This is an example of a source code with CWE-328 (Use of Weak Hash)~\cite{cwe_328}. 

\begin{listing}[!ht]
\begin{PythonSourceCode*}{firstnumber=10,highlightlines=24}
# ... prior lines hidden due to space constraints ...
@app.route('/create/user', methods=['POST'])
def create_user():
  data = request.get_json()
  username = data.get('username')
  email = data.get('email')
  password = data.get('password')
  # Check if all fields are provided
  if not all([username, email, password]):
    return jsonify({'message': 'Missing fields'}), 400 
  # Check if user already exists
  if User.query.filter_by(username=username).first():
    return jsonify({'message': 'Username not available'}), 400
  # Hash the password with MD5
  hashed_password = hashlib.md5(password.encode()).hexdigest()
  # Create and save the new user
  new_user = User(username, email, hashed_password)
  db.session.add(new_user)
  db.session.commit()
  return jsonify({'message': 'New user created'}), 201
\end{PythonSourceCode*}
\caption{Example of a Code Using Unsafe APIs (CWE-328)}\label{lst:cwe_328}
\end{listing}


To detect unsafe APIs being used in a generated code, our framework uses  CodeQL~\cite{codeql}. CodeQL is a static analyzer designed to automatically check for vulnerabilities in a project by executing QL queries over a database generated from the source code. CodeQL can be used to match the function of the function call. 


Besides unsafe API misuse,  several prompts in our database are related to injection vulnerabilities. These vulnerabilities are caused by \textit{untrusted data flows}~\cite{yamaguchi2015automatic, livshits2005finding}.  These vulnerabilities are traditionally detectable through \textit{taint analysis}, which is a technique that tracks flows of \textit{sources} of potentially untrusted (tainted) data (\eg parameters in HTTP requests) to sensitive program areas (\textit{sinks})~\cite{schwartz2010all}. 
%
%
%
%
In these cases, our framework uses CodeQL to perform (static) taint analysis of variables and check if they reach a sink method (\eg \snippetPython{os.system}).


\subsubsection{Evaluation Metrics}\label{subsec:SecurityChecker}

Models are commonly evaluated using the \texttt{pass@k} metric~\cite{chen2021codex, kulal2019spoc}. This metric evaluates the probability that \textit{at least one} out of $k$ generated samples are \textit{functionally correct} (\ie passed all \textit{functional} test cases). 
To evaluate the \texttt{pass@k}, we generate $n$ samples per prompt ($n \geq k$), count the number of  samples  $c$  that are functionally correct ($c \leq n$), and calculate the unbiased estimator  $\mathbb{E}$  by Kulal \etal \cite{kulal2019spoc}:

\begin{equation}
    pass@k = \mathbb{E}_{prompts}\left[1- \frac{\binom {n-c}k}{\binom nk} \right]
\end{equation}


Although the \texttt{pass@k} is a widely-used metric~\cite{kulal2019spoc,chen2021codex}, it does not measure the \textit{security} of the generated code.  Therefore, in this paper, we introduce two novel metrics (\texttt{secure@k} and \texttt{vulnerable@k}) for measuring the security of the generated code. These metrics are defined as follows:


\begin{itemize}[leftmargin=*]
    \item The \texttt{vulnerable@k} metric measures the probability that \textit{at least one} code snippet out of $k$ generated samples is vulnerable (\ie a vulnerability was detected by our assessment techniques). To compute this metric, we generate $n$ samples per prompt,  count the number  $v$  of generated vulnerable samples, and use the unbiased estimator in Eq.~\ref{eq:VulnAtK}. For this metric, {\textit{the model is better if the \texttt{vulnerable@k} score is lower}}.

\begin{equation}\label{eq:VulnAtK}
    vulnerable@k = \mathbb{E}_{prompts}\left[1- \frac{\binom {n-v}k}{\binom nk} \right]
\end{equation}

    \item The \texttt{secure@k} metric measures the probability that \textit{all} code snippets out of $k$ samples are vulnerability-free (\ie no vulnerability has been detected by our assessment techniques). That is,  the prompt is considered secure if \textit{all} of the generated code in the top-k passes our assessment techniques.  To clarify, consider that we have 10 prompts,  a model generates 10 outputs for each problem described in a prompt, and we sample 3 out of 10 outputs generated by the model. If our assessment technique does not detect any vulnerability in all the 3 sampled outputs for  6 prompts, then the \textit{secure@3} score will be 60\%.
    Threfore, to compute this metric, we count the number of prompts $s$  in which \textit{all} \textit{k} samples do not have a detected vulnerability in it and divided it by the number of prompts \textit{p}:

\begin{equation}
    secure@k = \frac{s}{p}
\end{equation}\label{eq:SecAtK}
\end{itemize}



It is important to highlight that our novel metrics (\texttt{secure@k} and \texttt{vulnerable@k}) can be computed statically, dynamically, or a combination of both. Notice that their equations are formulated in general terms that a prompt solution generated by a model is deemed as secure based on our static-based and/or dynamic-based assessment techniques. In our evaluation experiments (\secref{sec:RQ2Results}), we will demonstrate the computation of these metrics both statically (by using CodeQL) and dynamically (by leveraging unit tests).

\paragraph*{$\blacktriangleright$ Estimating the pass@k, and vulnerable@k}
Calculating Kulal \etal~\cite{kulal2019spoc} estimator directly results in large numbers and numerical instability~\cite{StarCoder}. Thus, to compute the \textit{pass@k}, and \textit{vulnerable@k} metrics, we used a numerically stable implementation from Chen \etal \cite{chen2021codex}.
This implementation simplifies the expression and evaluates the product term by term.

\section{Experiments}\label{sec:experiments}

This section describes the research questions we address in our experiments ($\S$~\ref{subsec:RQs}) as well as the methodology to answer each of these questions ($\S$~\ref{subsec:RQ1Methodology}--\ref{subsec:RQ2Methodology}).

\subsection{Research Questions}\label{subsec:RQs}

We aim to answer the following questions:

\begin{description}
    \item[\textbf{RQ1}]\textbf{ How does \sandoval's dataset of prompts  compare to existing datasets?}
\end{description}

First, we demonstrate the value
of our manually curated dataset of prompts by comparing it to two \textbf{peer-reviewed} datasets: LLMSecEval~\cite{llmseceval} and SecurityEval~\cite{siddiq2022seceval}. We contrast their coverage of vulnerability types (CWEs) and dataset size.

\begin{description}
    \item[RQ2] \textbf{How well do LLMs perform with security-centric prompts compared to the evaluation setting used in their original studies?}
    
\end{description}

As explained in ~\secref{subsec:Gaps}, LLMs are evaluated with respect to their ability to generate functional code (not necessarily secure). Thus, in this question, we  evaluate  the models' performance with respect to generating code that is both \textit{functionally correct} but also \textit{secure}.

    



\subsection{RQ1 Methodology}\label{subsec:RQ1Methodology}
To answer RQ1, we compare \sandoval's dataset to two  prior peer-reviewed datasets of prompts used to evaluate the security of LLM generated code:

\begin{itemize}[leftmargin=*, label=-,noitemsep,topsep=0pt]
    \item \textbf{SecurityEval}~\cite{siddiq2022seceval} is a prompt-based dataset covering 69 CWEs, including the MITRE’s Top 25 CWEs. The prompts are signatures of Python functions along with their docstrings and import statements.
    \item \textbf{LLMSecEval}~\cite{llmseceval} is a natural language (NL) prompt-to-code dataset crafted from Pearce \etal \cite{pearce2021}. 
\end{itemize}

We compare these datasets according to two dimensions:
\textsc{(i)} {\textit{number of supported vulnerability types (CWEs)}};  \textsc{(ii)} {\textit{dataset size (number of prompts)}} and \textsc{(iii)} {\textit{prompt style}}.

\subsection{RQ2 Methodology}\label{subsec:RQ2Methodology}

We investigate in RQ2 the performance of existing LLMs when evaluated using \sandoval, our framework. To answer this question, we provide each prompt in our dataset as inputs  to four models from three LLM families: 

\begin{itemize}[leftmargin=*, label=-,noitemsep,topsep=0pt]
    \item \textbf{\textsc{CodeGen}} ~\cite{Nijkamp2022ACP} is an LLM for code generation trained on three large code datasets. This model has three variants: \textsc{CodeGen-nl}, \textsc{CodeGen-multi}, and \textsc{CodeGen-mono}. 
    \textsc{CodeGen-nl} is trained with the \textit{Pile} dataset~\cite{ThePileDataset} is focused on text generation. The \textsc{CodeGen-multi} is built on top of \textsc{CodeGen-nl} but further trained with a large scale-dataset of code snippets in six different languages (\ie C, C++, Go, Java, JavaScript, and Python)~\cite{BigQueryDataset}. The \textsc{CodeGen-mono} is built from \textsc{CodeGen-multi} and further trained with a dataset~\cite{Nijkamp2022ACP} of only Python code snippets. They also released another version called \textsc{CodeGen2.5} \cite{nijkamp2023codegen2} which is trained on the StarCoder data from BigCode \cite{Kocetkov2022TheStack}. It has a mono and multi version.
    Since the latter variant is focused on Python-only generation,  we use \textbf{\textsc{CodeGen-2B-mono}} and \textbf{\textsc{CodeGen-2.5-7B-mono}} to generate Python code.

    \item \textbf{\textsc{StarCoder}}~\cite{StarCoder}  is an LLM with 15.5B parameters trained with over 80 different programming languages. This model is focused on fill-in-the-middle objectives and can complete code given a code-based prompt. 
    \item The \textbf{\textsc{Generative Pre-trained Model (GPT)}} ~\cite{brown2020language} is a family of  transformer-based~\cite{attention2017} and task-agnostic models capable of  generating source code. We used the latest OpenAI's GPT models, \ie \textbf{\textsc{GPT-3.5-Turbo}} and \textbf{\textsc{GPT-4}},  which are tuned for chat-style conversation and powers a popular chat-based question-answering tool, ChatGPT \cite{chatgpt} and its paid variant (ChatGPT plus).
\end{itemize}

We chose these models based on their availability and performance from a leaderboard using the most commonly used benchmark, HumanEval \cite{humanevalleaderboard} when this study was conducted and because prior works~\cite{openai2023gpt4, nijkamp2023codegen2,siddiq2022seceval,siddiq2024regexeval,siddiq2023franc,siddiq2024devgpt}  have studied their performance. Most of the top models are a variation of GPT models. We also used three top open-source models.

We configure each LLM to generate \textbf{10} code solutions for each prompt with \textbf{256} new tokens. We selected 256 as the token size to generate because we observed that the insecure code examples in our dataset have an average of 54 tokens and a maximum of 245 tokens. Thus, a 256 token size would be sufficient for the models. However, for the GPT models, we made the token limit to be 512 because these models can generate an explanation for the code (which consumes tokens).

Furthermore,  we vary the models'  \textit{temperature} parameter from 0 to 1 in 0.2 increments (\ie 0.0, 0.2, 0.4, 0.6, 0.8, and 1.0). This way we can observe the performance across different \textit{temperatures}, which is a parameter that controls the randomness of the model's generations (lower temperature values yield more predictable and repetitive outputs).

After obtaining the generated code solutions from each model, we measure and contrast the performance of these models with respect to three metrics: \textit{pass@k}~\cite{chen2021codex}, \textit{vulnerable@k}
and \textit{secure@k} (the last two are our novel metrics, as defined in ~\secref{subsec:SecurityChecker}). 
In our experiments, we chose $k$ to be equal to 1, 3, and 5. 
This is because our goal is to evaluate these models for typical use scenarios, where developers will likely inspect only the first few generated code snippets by a model.

\section{Results}\label{sec:results}

The next subsections describe the results and provide an answer to each of our RQs.

\subsection{RQ1 Results}

Table~\ref{tab:rq1-results} contrasts each dataset, including our framework's dataset (denoted by \sandoval{} on this table).

\begin{table}[!h]
\caption{Dataset comparison}
\scriptsize\centering
\label{tab:rq1-results}
    \setlength\tabcolsep{4.9pt} 
\begin{tabular}{@{}rccccc@{}}
\toprule
\multicolumn{1}{c}{\textbf{Datasets}} & \textbf{\# Prompts} & \textbf{\begin{tabular}[c]{@{}c@{}}\# Python\\ Prompts\end{tabular}} & \textbf{\# CWEs} & \textbf{\begin{tabular}[c]{@{}c@{}}Prompt\\ Style\end{tabular}} & \textbf{Language(s)} \\ \midrule
\textbf{LLMSecEval} & 150 & 83 & 18 & Text & C and Python \\
\textbf{SecurityEval} & 121 & 121 & 69 & Code & Python \\
\textbf{\sandoval} & 100 & 100 & 45 & Code and Text & Python \\ \bottomrule
\end{tabular}
\end{table}

\subsubsection{CWE Coverage}

As shown in this table, our dataset covers 2.5 times more CWEs (45 CWEs) than LLMSecEval \cite{llmseceval}, which covers only 18 CWEs (a subset of the CWE top 25~\cite{top25}). In contrast, SecurityEval \cite{siddiq2022seceval} covers 69 CWEs, whereas \sandoval's dataset has a slightly less amount of CWEs. 

Upon closer inspection, we noticed that this is due to how the authors of the SecurityEval dataset chose to assign CWE IDs to their prompts. 
The CWE list includes hierarchical relationships (\eg \textit{CWE-89: SQL Injection} is a child of \textit{CWE-943:	Improper Neutralization of Special Elements in Data Query Logic}).
In our dataset, we follow MITRE's CWE mapping~\cite{CWEMappingGuidance} to \textit{consistently} map prompts to CWE IDs that were at the lowest level possible of the CWE hierarchy (\ie as more specialized as possible). SecurityEval, on the other hand, has some prompts tagged with higher-level abstraction CWEs and other with more specific ones. This inconsistency increases the number of CWEs.

\subsubsection{Dataset Size}

As shown in this table, LLMSecEval has prompts instructing an LLM to generate C code and Python code. Out of their 150 prompts, only 83 of them are for Python. SecurityEval has a total of 121 prompts. It is important to highlight that although SecurityEval has more prompts than \sandoval's dataset, its dataset size in terms of number of tokens is \textit{smaller} than ours. \sandoval's dataset prompts have an average of  265 tokens, whereas SecurityEval's prompts have 157 tokens on average.
Moreover, we also found several SecurityEval's  prompts that were not compilable because they required external libraries or were single scripts that are meant to be part of a codebase (\eg a Django application) and these other parts were missing.

\subsubsection{Prompt Style}
LLMSecEval's prompts are natural language prompts in the form of \textit{``Generate [language] code for the following: [coding problem description]''}. Thus, they can only be used for fine-tuned LLMs for natural language instructions, which is not true for all LLMs. For example, StarCoder~\cite{StarCoder} is an LLM that was not trained for natural language prompts\footnote{As described in StarCode's intended use section~\cite{StarCoderIntendedUsage}: \textit{``[StarCoder] was trained on GitHub code. As such, it is not an instruction model, and commands like "Write a function that computes the square root." do not work well.''}} and, as a result, is unable to understand prompts in the form of \textit{"Write a Python code that parses a CSV file."}. SecurityEval is the opposite: it only includes prompts as a docstring in a function to be completed. 
Unlike these datasets, \sandoval{} includes prompts in both styles, allowing it to be used by models that can understand only text or that can understand only code.

It is also important to highlight that LLMSecEval~\cite{llmseceval} and SecurityEval~\cite{siddiq2022seceval} do not include an automated execution environment for evaluation. As such, these datasets do not provide a necessary automation to enable a systematic and automated benchmarking of models. Unlike these works, each prompt in \sandoval's dataset contains runnable test cases to test a generated code's correctness and security as well as an execution environment.




\rques{
\textbf{RQ1 Summary of Findings}:
\begin{itemize}[leftmargin=*]
    \item \sandoval's dataset has \nPrompts{} Python prompts that are suitable for  models that can understand code and/or text. Our dataset covers 45 vulnerability types (CWEs). 
    \item \sandoval's prompts is both in code format and  textual format, which makes it suitable for models that accept code-only or text-only prompts. 
    \item Unlike LLMSecEval and SecurityEval, all \sandoval's prompts include a runnable insecure solution example, a set of test cases, and a docker environment that enables automated and systematic output evaluation of models.
\end{itemize}

}

\subsection{RQ2 Results}\label{sec:RQ2Results}

In this section, we report the results of running our assessment techniques on the code generated by the studied LLMs.

\subsubsection{Syntactic Correctness}
\begin{figure}[!ht]
    \centering
    \includegraphics[width=\linewidth]{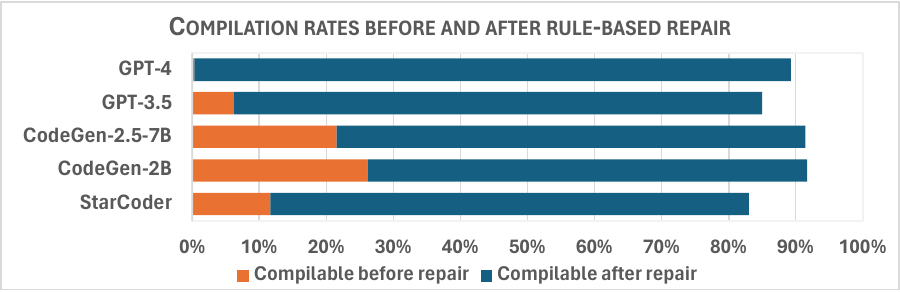}
    \caption{Compilation rates before and after applying \sandoval's repair rules}
    \label{fig:Compilation}
\end{figure}

As described in ~\secref{subsec:CodeGeneration}, \sandoval{} includes a rule-based repair component to automatically fix common compilations issues that models produce~\cite{siddiq2023exploring}. \figref{fig:Compilation} depicts the percentage of code snippets that were compilable to Python bytecode \textit{before} and \textit{after} \sandoval's repair step. Our framework increases the compilation rates of generated code from \textbf{15\%} to \textbf{75\%}, on average. The model that \sandoval{} repaired the most was GPT-4; increasing its compilation rates from less than \textbf{1\%} to \textbf{89\%}. \sandoval's rule \textbf{R1}, which removes natural text from the model's output, was the most used rule to repair scripts generated by GPT-4.

\subsubsection{Functional Correctness (pass@k metric)}\label{subsec:RQ2PassAtK}

\tblref{tab:rq2_pass_at_k} contains the results for the \texttt{pass@k} for each studied LLM and temperature combination. The numbers in \green{dark green} are those that had the \textit{best} performance for a given metric; the numbers in \red{dark red} are those in which the model had the \textit{worst} performance.  

\begin{table}[!ht]
\centering\scriptsize
\setlength\tabcolsep{2.5pt} 
\caption{\texttt{pass@k} for different models and temperatures.}
\label{tab:rq2_pass_at_k}
\begin{tabular}{ccccccc}
\toprule
\textbf{Temp}                   & \textbf{Metric} & \textbf{CodeGen-2B} & \textbf{CodeGen-2.5-7B} & \textbf{StarCoder} & \textbf{GPT-3.5} & \textbf{GPT-4} \\ \midrule
\multirow{3}{*}{\textbf{0.0}}   & \textbf{pass@1} & \red{28} & - & - & 42 & \green{48.8} \\
                                & \textbf{pass@3} & \red{28} & - & - & 42 & \green{49} \\
                                & \textbf{pass@5} & \red{28} & - & - & 42 & \green{49} \\\hline
\multirow{3}{*}{\textbf{0.2}}   & \textbf{pass@1} & 24.9 & 37.4 & \red{8.0} & 41.2 & \green{46.4} \\
                                & \textbf{pass@3} & 33.8 & 45.4 & \red{15.2} & 44.9 & \green{49.5} \\
                                & \textbf{pass@5} & 37.5 & 47.7 & \red{17.9} & 46.0 & \green{50.4} \\\hline
\multirow{3}{*}{\textbf{0.4}}   & \textbf{pass@1} & 24.2 & 38.1 & \red{9.2} & 39.6 & \green{47.1} \\
                                 & \textbf{pass@3} & 37.8 & 48.7 & \red{19.0} & 46.5 & \green{52.1} \\
                                 & \textbf{pass@5} & 43.3 & 52.4 & \red{24.8} & 48.0 & \green{53.3} \\ \hline
\multirow{3}{*}{\textbf{0.6}}    & \textbf{pass@1} & 21.3 & 34.3 & \red{8.6} & 40.0 & \green{44.5} \\
                                 & \textbf{pass@3} & 36.0 & 49.9 & \red{18.7} & 50.2 & \green{52.0} \\
                                 & \textbf{pass@5} & 42.3 & \green{54.6} & \red{25.1} & 53.4 & 53.7 \\ \hline
\multirow{3}{*}{\textbf{0.8}}   & \textbf{pass@1} & 17.1 & {26.3} & \red{6.7} & \green{37.6} & {20.7} \\
                                 & \textbf{pass@3} & 33.4 & {45.1} & \red{16.2} & \green{50.4} & 23.5 \\
                                 & \textbf{pass@5} & 41.7 & 52.2 & \red{22.6} & \green{54.3} & 23.9 \\ \hline
\multirow{3}{*}{\textbf{1.0}} & \textbf{pass@1} & 10.0 & 16.7 & \red{5.5} & 35.8 & \green{42.1} \\
                                 & \textbf{pass@3} & 21.8 & 34.6 & \red{14.3} & 49.7 & \green{50.9} \\
                                 & \textbf{pass@5} & 28.8 & 43.6 & \red{20.9} & \green{54.8} & 52.9 \\ \bottomrule
\end{tabular}
\end{table}

The \texttt{pass@1}, \texttt{pass@3}, and \texttt{pass@5} across all models ranged from 5.5\% to 54.8\%. GPT-4 consistently outperformed the remaining models for the temperatures 0, 0.2, and 0.4. For higher temperatures, the best performing model included not only GPT-4, but also its older version (GPT-3.5), and CodeGen-2.5-7B.
StarCoder was the worst performing model with respect to generating functionally correct code. Its \texttt{pass@k} was an average of 15.5\% (ranging from 5.5\% to 25.1\%).

\subsubsection{Security (secure@k and vulnerable@k metrics)}
We compute the \texttt{secure@k} and \texttt{vulnerable@k} metrics based on the \textbf{\textit{static-based assessment}} technique and the \textbf{\textit{test-based assessment}} technique. These results are discussed in the next paragraphs.

\paragraph{Static-based Assessment}
\tblref{tab:rq2_all} presents the \texttt{vulnerable@k} and \texttt{secure@k}  computed based on the outcomes from \sandoval's \textit{\textbf{static-based}} assessment technique.  The \texttt{vulnerable@k} varied from 16\% to 59\%. For temperature 0, all models had the same \texttt{vulnerable@1}, \texttt{vulnerable@3}, and \texttt{vulnerable@5} as well as their \texttt{secure@1}, \texttt{secure@3}, and \texttt{secure@5}. This is caused by the fact that the temperature 0 makes the results more \textit{predictable}, \ie the generated output has less variance.

From these results, we observe that StarCoder  had the lowest \texttt{vulnerable@k} across all temperatures. On the other hand, CodeGen-2B and CodeGen-2.5-7B had a worse performance, on average, than the other LLMs. For the GPT-style models, GPT-4 performed better than GPT-3.5-Turbo.

\begin{table*}[!ht]
\caption{Static Analysis-based and Test-based computation of \texttt{secure@k} and \texttt{vulnerable@k} for different models.}\scriptsize
\setlength\tabcolsep{2.8pt} 
\label{tab:rq2_all}
\begin{tabular}{ccccccccccccccccc}
\toprule
 &  & \multicolumn{3}{c}{\textbf{CodeGen-2B}} & \multicolumn{3}{c}{\textbf{CodeGen-2.5-7B}} & \multicolumn{3}{c}{\textbf{StarCoder}} & \multicolumn{3}{c}{\textbf{GPT-3.5}} & \multicolumn{3}{c}{\textbf{GPT-4}} \\
\textbf{Temperature} & \textbf{Metrics} & \textbf{\begin{tabular}[c]{@{}c@{}}Test-\\ based\end{tabular}} & \textbf{\begin{tabular}[c]{@{}c@{}}Static-\\ based\end{tabular}} & \textbf{\begin{tabular}[c]{@{}c@{}}Harmonic\\ Mean\end{tabular}} & \textbf{\begin{tabular}[c]{@{}c@{}}Test-\\ based\end{tabular}} & \textbf{\begin{tabular}[c]{@{}c@{}}Static-\\based\end{tabular}} & \textbf{\begin{tabular}[c]{@{}c@{}}Harmonic\\ Mean\end{tabular}} & \textbf{\begin{tabular}[c]{@{}c@{}}Test-\\ based\end{tabular}} & \textbf{\begin{tabular}[c]{@{}c@{}}Static-\\based\end{tabular}} & \textbf{\begin{tabular}[c]{@{}c@{}}Harmonic\\ Mean\end{tabular}} & \textbf{\begin{tabular}[c]{@{}c@{}}Test-\\ based\end{tabular}} & \textbf{\begin{tabular}[c]{@{}c@{}}Static-\\based\end{tabular}} & \textbf{\begin{tabular}[c]{@{}c@{}}Harmonic\\ Mean\end{tabular}} & \textbf{\begin{tabular}[c]{@{}c@{}}Test-\\ based\end{tabular}} & \textbf{\begin{tabular}[c]{@{}c@{}}Static-\\based\end{tabular}} & \textbf{\begin{tabular}[c]{@{}c@{}}Harmonic\\ Mean\end{tabular}} \\\midrule
\multirow{6}{*}{\textbf{0.0}} & \textbf{vulnerable@1} & 50 & 38 & \green{43.2} & - & - & - & - & - & - & 49 & 51 & 50.0 & 52.7 & 48 & \red{50.2} \\
 & \textbf{vulnerable@3} & 50 & 38 & \green{43.2} & - & - & - & - & - & - & 49 & 51 & 50.0 & 53.0 & 48 & \red{50.4} \\
 & \textbf{vulnerable@5} & 50 & 38 & \green{43.2} & - & - & - & - & - & - & 49 & 51 & 50.0 & 53 & 48 & \red{50.4} \\
 & \textbf{secure@1} & 23 & 62 & \green{33.6} & - & - & - & - & - & - & 22 & 49 & 30.4 & 17 & 52 & \red{25.6} \\
 & \textbf{secure@3} & 22 & 62 & \green{32.5} & - & - & - & - & - & - & 22 & 49 & 30.4 & 17 & 52 & \red{25.6} \\
 & \textbf{secure@5} & 21 & 62 & \green{31.4} & - & - & - & - & - & - & 22 & 49 & 30.4 & 17 & 52 & \red{25.6} \\\hline
\multirow{6}{*}{\textbf{0.2}} & \textbf{vulnerable@1} & 46.6 & 39.7 & 42.9 & 55.4 & 46.4 & \red{50.5} & 37.8 & 19.8 & \green{26.0} & 47.2 & 49.5 & 48.3 & 50.4 & 47.1 & 48.7 \\
 & \textbf{vulnerable@3} & 53.6 & 46.8 & 50.0 & 61.2 & 50.7 & \red{55.4} & 46.4 & 27.6 & \green{34.6} & 50.8 & 50.8 & 50.8 & 54.2 & 47.8 & 50.8 \\
 & \textbf{vulnerable@5} & 55.9 & 48.8 & 52.1 & 63.1 & 51.7 & \red{56.8} & 49.3 & 30.3 & \green{37.5} & 52.2 & 51.0 & 51.6 & 55.2 & 50.0 & 52.5 \\
 & \textbf{secure@1} & 24.0 & 61.0 & \green{34.4} & 25.0 & 51.0 & 33.6 & 19.0 & 82.0 & 30.9 & 22.0 & 49.0 & 30.4 & 19.0 & 52.0 & \red{27.8} \\
 & \textbf{secure@3} & 20.0 & 51.0 & \green{28.7} & 18.0 & 47.0 & 26.0 & 11.0 & 74.0 & \red{19.2} & 17.0 & 49.0 & 25.2 & 14.0 & 52.0 & 22.1 \\
 & \textbf{secure@5} & 17.0 & 50.0 & \green{25.4} & 17.0 & 47.0 & 25.0 & 10.0 & 67.0 & \red{17.4} & 14.0 & 49.0 & 21.8 & 13.0 & 52.0 & 20.8 \\\hline
\multirow{6}{*}{\textbf{0.4}} & \textbf{vulnerable@1} & 46.3 & 40.1 & 43.0 & 53.7 & 44.7 & 48.8 & 35.9 & 18.9 & \green{24.8} & 45.8 & 47.8 & 46.8 & 52.4 & 46.7 & \red{49.4} \\
 & \textbf{vulnerable@3} & 58.3 & 49.6 & 53.6 & 62.8 & 51.5 & \red{56.6} & 49.2 & 30.0 & \green{37.3} & 51.4 & 51.2 & 51.3 & 56.4 & 48.0 & 51.8 \\
 & \textbf{vulnerable@5} & 61.8 & 53.1 & 57.1 & 64.9 & 52.9 & \red{58.3} & 53.6 & 35.0 & \green{42.4} & 53.4 & 52.0 & 52.7 & 57.6 & 48.9 & 52.9 \\
 & \textbf{secure@1} & 22.0 & 59.0 & 32.0 & 24.0 & 55.0 & \green{33.4} & 18.0 & 79.0 & 29.3 & 22.0 & 53.0 & 31.1 & 18.0 & 52.0 & \red{26.7} \\
 & \textbf{secure@3} & 17.0 & 49.0 & 25.2 & 18.0 & 51.0 &  \green{26.6} & 10.0 & 70.0 & \red{17.5} & 13.0 & 50.0 & 20.6 & 14.0 & 52.0 & 22.1 \\
 & \textbf{secure@5} & 16.0 & 42.0 &  \green{23.2} & 15.0 & 46.0 & 22.6 & 8.0 & 57.0 & 14.0 & 8.0 & 47.0 & \red{13.7} & 10.0 & 51.0 & 16.7 \\\hline
\multirow{6}{*}{\textbf{0.6}} & \textbf{vulnerable@1} & 44.1 & 37.1 & 40.3 & 51.3 & 43.3 & 47.0 & 34.1 & 20.2 &  \green{25.4} & 46.3 & 46.2 & 46.2 & 49.7 & 45.9 & \red{47.7} \\
 & \textbf{vulnerable@3} & 59.0 & 50.6 & 54.5 & 61.5 & 53.2 & \red{57.1} & 49.3 & 35.2 &  \green{41.1} & 54.7 & 51.2 & 52.9 & 55.0 & 47.8 & 51.1 \\
 & \textbf{vulnerable@5} & 63.0 & 54.1 & 58.2 & 63.5 & 57.0 & \red{60.1} & 55.3 & 41.6 &  \green{47.5} & 57.8 & 52.4 & 55.0 & 56.0 & 48.0 & 51.7 \\
 & \textbf{secure@1} & 20.0 & 60.0 & 30.0 & 29.0 & 53.0 &  \green{37.5} & 21.0 & 83.0 & 33.5 & 27.0 & 53.0 & 35.8 & 12.0 & 53.0 & \red{19.6} \\
 & \textbf{secure@3} & 13.0 & 52.0 & 20.8 & 20.0 & 41.0 &  \green{26.9} & 9.0 & 71.0 & 16.0 & 11.0 & 47.0 & 17.8 & 8.0 & 52.0 & \red{13.9} \\
 & \textbf{secure@5} & 12.0 & 43.0 &  \green{18.8} & 12.0 & 38.0 & 18.2 & 6.0 & 52.0 & \red{10.8} & 9.0 & 47.0 & 15.1 & 7.0 & 52.0 & 12.3 \\\hline
\multirow{6}{*}{\textbf{0.8}} & \textbf{vulnerable@1} & 41.3 & 34.3 & 37.5 & 45.3 & 36.6 & 40.5 & 31.5 & 19.0 &  \green{23.7} & 45.6 & 47.2 & \red{46.4} & 16.8 & 43.9 & 24.3 \\
 & \textbf{vulnerable@3} & 58.4 & 50.8 & 54.3 & 62.7 & 51.3 & \red{56.4} & 48.7 & 34.4 & 40.3 & 56.0 & 52.2 & 54.1 & 20.3 & 48.3 &  \green{28.6} \\
 & \textbf{vulnerable@5} & 63.0 & 55.3 & 58.9 & 67.7 & 55.8 & \red{61.2} & 55.4 & 41.2 & 47.2 & 60.3 & 53.4 & 56.7 & 21.0 & 49.7 &  \green{29.5} \\
 & \textbf{secure@1} & 12.0 & 65.0 & \red{20.3} & 27.0 & 69.0 &  \green{38.8} & 21.0 & 77.0 & 33.0 & 26.0 & 57.0 & 35.7 & 19.0 & 56.0 & 28.4 \\
 & \textbf{secure@3} & 10.0 & 50.0 & 16.7 & 13.0 & 52.0 &  \green{20.8} & 7.0 & 62.0 & \red{12.6} & 10.0 & 50.0 & 16.7 & 9.0 & 52.0 & 15.3 \\
 & \textbf{secure@5} & 6.0 & 41.0 & 10.5 & 8.0 & 39.0 &  \green{13.3} & 4.0 & 50.0 & \red{7.4} & 7.0 & 45.0 & 12.1 & 4.0 & 48.0 & 7.4 \\\hline
\multirow{6}{*}{\textbf{1.0}} & \textbf{vulnerable@1} & 37.7 & 30.0 & 33.4 & 36.6 & 31.5 & 33.9 & 27.6 & 16.3 &  \green{20.5} & 44.8 & 44.2 & 44.5 & 47.8 & 43.9 & \red{45.8} \\
 & \textbf{vulnerable@3} & 56.8 & 47.7 & 51.9 & 57.7 & 52.0 & \red{54.7} & 45.9 & 31.7 &  \green{37.5} & 56.6 & 51.2 & 53.8 & 56.7 & 48.3 & 52.2 \\
 & \textbf{vulnerable@5} & 62.3 & 52.6 & 57.0 & 64.1 & 59.1 & \red{61.5} & 53.7 & 39.6 &  \green{45.6} & 60.2 & 53.6 & 56.7 & 58.7 & 49.7 & 53.8 \\
 & \textbf{secure@1} & 18.0 & 68.0 & 28.5 & 21.0 & 64.0 & 31.6 & 18.0 & 82.0 & 29.5 & 26.0 & 56.0 &  \green{35.5} & 12.0 & 56.0 & \red{19.8} \\
 & \textbf{secure@3} & 11.0 & 56.0 &  \green{18.4} & 7.0 & 48.0 & 12.2 & 3.0 & 68.0 & \red{5.7} & 11.0 & 48.0 & 17.9 & 8.0 & 52.0 & 13.9 \\
 & \textbf{secure@5} & 10.0 & 44.0 &  \green{16.3} & 4.0 & 35.0 & 7.2 & 2.0 & 50.0 & \red{3.8} & 4.0 & 43.0 & 7.3 & 6.0 & 48.0 & 10.7 \\ \bottomrule
\end{tabular}
\end{table*}

\paragraph{Test-based Assessment}

\tblref{tab:rq2_all} shows  the \texttt{vulnerable@k} and \texttt{secure@k}  computed based on \sandoval's \textbf{\textit{test-based}} assessment technique.
The models had similar performance with respect to \texttt{secure@k}, with GPT-3.5 and CodeGen-2B performing slightly better, on average. For \texttt{vulnerable@k}, StarCoder, on average, performed better than the other models. This result is consistent to what was observed on the static-based assessment of these metrics.

\tblref{tab:rq2_all} reports the harmonic mean between the secure@k and vulnerable@k when computed using static-based and test-based assessment techniques.
We use \green{dark green} and \red{dark red}  to flag the \textit{best} and \textit{worst} performance for a given metric, respectively. Recall that for the \texttt{vulnerable@k} metric, a \textit{lower} value is better.
When looking at the combined performance of models for these two different assessment techniques, we observe that there is not a clear model that consistently outperforms the other across all temperatures.

\subsubsection{Overall Performance}

To better understand the models' performance with respect to being able to generate code that is \textbf{\textit{both}}
\textit{functionally correct} and \textit{secure}, we computed the harmonic mean of the \textit{pass@k} and \textit{secure@k}. The secure@k is computed as the harmonic mean of the secure@k computed via tests and via static analysis. 
These results are presented in Table~\ref{tab:rq2_overall}.

\begin{table}[!ht]
\centering\scriptsize
\setlength\tabcolsep{2.5pt} 
\caption{The harmonic mean for the \texttt{pass@k} and \texttt{secure@k} for different models and temperatures. 
}
\label{tab:rq2_overall}
\begin{tabular}{ccccccc}
\toprule
\textbf{Temp}                   & \textbf{Metric} & \textbf{CodeGen-2B} & \textbf{CodeGen-2.5-7B} & \textbf{StarCoder} & \textbf{GPT-3.5} & \textbf{GPT-4} \\ \midrule
 & harmonic mean@1 & \red{30.5} & - & - & \green{35.3} & 33.6 \\
 & harmonic mean@3 & \red{30.1} & - & - & \green{35.3} & 33.6 \\
\multirow{-3}{*}{0} & harmonic mean@5 & \red{29.6} & - & - & \green{35.3} & 33.6 \\\hline
 & harmonic mean@1 & 28.9 & \green{35.4} & \red{12.7} & 35.0 & 34.8 \\
 & harmonic mean@3 & 31.1 & \green{33.1} & \red{17.0} & 32.3 & 30.6 \\
\multirow{-3}{*}{0.2} & harmonic mean@5 & 30.3 & \green{32.8} & \red{17.6} & 29.6 & 29.5 \\\hline
 & harmonic mean@1 & 27.6 & \green{35.6} & \red{14.0} & 34.8 & 34.1 \\
 & harmonic mean@3 & 30.3 & \green{34.4} & \red{18.2} & 28.5 & 31.0 \\
\multirow{-3}{*}{0.4} & harmonic mean@5 & 30.2 & \green{31.6} & \red{17.9} & 21.3 & 25.4 \\\hline
 & harmonic mean@1 & 24.9 & 35.8 & \red{13.7} & \green{37.8} & 27.2 \\
 & harmonic mean@3 & 26.4 & \green{35.0} & \red{17.2} & 26.3 & 21.9 \\
\multirow{-3}{*}{0.6} & harmonic mean@5 & 26.0 & \green{27.3} & \red{15.1} & 23.5 & 20.0 \\\hline
 & harmonic mean@1 & 18.6 & 31.3 & \red{11.1} & \green{36.6} & 23.9 \\
 & harmonic mean@3 & 22.3 & \green{28.5} & \red{14.2} & 25.1 & 18.5 \\
\multirow{-3}{*}{0.8} & harmonic mean@5 & 16.8 & \green{21.2} & \red{11.1} & 19.8 & 11.3 \\\hline
 & harmonic mean@1 & 14.8 & 21.9 & \red{9.3} & \green{35.6} & 26.9 \\
 & harmonic mean@3 & 20.0 & 18.0 & \red{8.1} & \green{26.3} & 21.8 \\
\multirow{-3}{*}{1} & harmonic mean@5 & 20.8 & 12.4 & \red{6.4} & \green{12.9} & 17.8 \\ \bottomrule
\end{tabular}
\end{table}

These results show that, on one hand, CodeGen-2.5-7B was the model that struck a better balance between functional correctness and security. On the other hand, we also found that while GPT-4 was the best model in generating functionally correct code ($\S$~\ref{subsec:RQ2PassAtK}), it does not perform as well in generating secure code. Surprisingly, its older version (GPT-3.5) performed better at balancing correctness and security.

\rques{
\textbf{RQ2 Findings}:
\begin{itemize}[leftmargin=*]
    \item StarCoder generated more secure code than CodeGen-2B, CodeGen-2.5-7B, GPT-3.5, and GPT-4 from the perspective of \texttt{vulnerable@k}.
    \item CodeGen-2.5-7B was the model that struck a better balance between functional correctness and security.
\end{itemize}
}
\section{Discussion}
\label{sec:discussion}

Along with the two RQs answered in this work, we also identified  important implications for researchers and practitioners  as follows:

\begin{itemize}[leftmargin=*]
    \item \textbf{Co-relation between Functional Correctness and Security}
LLMs should generate not only functional code but also secure code so that they don't introduce vulnerabilities when integrated into a system's code base. The evaluation results discussed  in Section \ref{sec:RQ2Results} showed that  GPT models perform better in generating \textit{functionally correct} code. It is also noticeable that an open-source model, CodeGen-2.5, has a comparable result with respect to these closed-source LLMs.  If we compare the \texttt{vulnerable@k}, we can see that except for temperature 0.8, StarCoder is generating less vulnerable codes, but it was the worst model for generating function correct code. From the perspective of \texttt{secure@k}, we can see that for GPT-4, \texttt{secure@1} is the highest for most of the temperatures (\ie~except for temperature 0.8). This indicates that the first code generated by this model is usually vulnerable. If we consider \texttt{secure@5} (\ie~all 5 of the generated codes are secure), we can see that for most of the temperatures (\ie~except for temperature 0.4), StarCoder has the worst performance. Hence, our framework provides multiple perspectives around functional correctness and security, and it implies a trade-off for choosing the right model. For example, if we focus mostly on functional correctness, the GPT-4 model is the best option, but for most of the cases, its first generated code is not secure.

\item \textbf{Implication for the Developers and Researchers}
Developers are adopting LLMs for software engineering tasks, but to choose an appropriate model, they have to consider the privacy of their data, the accuracy of the generated code, and security. Open-source models can provide privacy of the data, as they are not shared with the closed model with APIs. However, according to our results in Section \ref{sec:RQ2Results}, the correctness of the generated code from open-source models is not comparatively better than GPT models (closed-source models), but the CodeGen-2.5 with 7 billion parameter model can have a close performance. 

With our framework, developers can automatically benchmark their set of model choices. Our framework includes a rule-based repair part, which can significantly increase the compilation rate of the generated code. 

While our work introduce two novel security-centric metrics, there is still a need for researchers to work on other quality attributes of the code. 
We need to benchmark which model can produce not only functionally correct, and secure code but also that fulfill other quality attributes, such as performance.
\end{itemize}

\subsection{Limitations and Threats to the Validity }\label{subsec:Limitations}

\sandoval's dataset contains only Python prompts, which is a generalizability threat to this work. However, Python is not only a popular language among developers~\cite{StackOverflowDevSurvey} but also a language that tends to be the one chosen for evaluation, as HumanEval~\cite{chen2021codex} is a dataset of Python-only prompts.

A threat to the internal validity of this work is the fact that the prompts were manually created from examples obtained from several sources (\eg CWE list). 
However, these prompts were created by two of the authors, one with over 10 years of programming experience  and the other with over 3 years of programming experience. 
We also conducted a peer review of the prompts to ensure their quality and clarity.

We used GitHub's CodeQL \cite{codeql} as a static analysis to measure the vulnerability of code samples. As this is a static analyzer, one threat to our work is that it can suffer from imprecision. However, it is important to highlight that our framework evaluates code samples from two perspectives: static-based and dynamic-based (via tests). These approaches are complementary and help mitigate this threat. 

\section{Related Work}\label{sec:relatedWork}


\subsection{Empirical Studies of Code Generation Models}

Automated code generation techniques were initially focused on deducting the users' intent from a high-level specification or input-output examples \cite{gulwani2017program,green69, manna2017}. These approaches transform task specifications into constraints, and the program is extracted after demonstrating its ability to satisfy the constraints \cite{gulwani2017program}.
With the rise of attention-based transformer models \cite{attention2017}, code generation has been treated as a sequence-to-sequence problem where the user intent comes in the form of natural language. Many LLMs have been produced to generate code, such as CodeBert \cite{codebert}, Codex \cite{chen2021codex}, and CodeT5 \cite{codet5}. 

Though the performance of the code generation task is increasing daily and user end tools like GitHub Copilot are being adapted by users \cite{shani2023survey}, they are not free of security issues. Pearce \etal \cite{pearce2021} studied the output of GitHub Copilot with their early release. They found that 40\% of the outputs are vulnerable. Siddiq \etal \cite{siddiq2022empirical} explored the code generative models and their datasets by following standard coding practices and security issues. Sandoval \etal \cite{sandoval2022security} measured if an AI assistant generates more vulnerable codes than users. Siddiq \etal \cite{siddiq2023franc} suggested a static analyzer-based ranking system to have more secured code in the output. Hajipour \etal \cite{hajipour2023systematically} investigated finding the vulnerabilities in the black box code generation model.

While there is a recent growing body of peer-reviewed literature that investigated the capabilities of code generation beyond their functional correctness but also security~\cite{dakhel2023github,sobania2022choose,nguyen2022empirical,pearce2021,sandoval2022security,perry2022users},  these existing studies only pinpoint the observed issues without proposing new metrics or a way to systematically benchmarking LLMs with respect to the security of the LLM generated code. Unlike these previous studies, in this paper, we release a dataset and an evaluation environment that can automatically benchmark code LLMs with respect to security.

\subsection{Benchmarks for Code LLMs}

Traditionally, deep learning models use a training set for learning and a test set to evaluate the model. For example, CodeXGlue \cite{codexglue} includes the Concode dataset \cite{iyer2018mapping}  for Java code generation, which contains a test set of 2,000 samples. 

The authors of the Codex \cite{chen2021codex} model developed HumanEval for this purpose. HumanEval contains 164 simple programming problems with canonical solutions and test cases. Mostly Basic Python Problems Dataset (MBPP) dataset contains around 1,000 samples for a similar purpose \cite{austin2021program}. These datasets are later extended for different programming languages \cite{zheng2023codegeex, MBXP}. CoderEval dataset \cite{yu2023codereval} uses samples from real-world software. However, these datasets focus on functional correctness. 

Pearce \etal \cite{pearce2021} provided a set of scenarios for testing the security of the generated code. SecurityEval \cite{siddiq2022seceval} formalized the prompts for testing security for many CWEs. Though these datasets focus on measuring security, they do not enable an automated and systematic approach for benchmarking LLMs provided by our framework. There are datasets for security evaluation from natural language prompts \cite{CodeLMSec}, but in their case, they only focus on finding the vulnerabilities in the generated code, not focusing on the functionality, whereas our focus is on both perspectives. 
The meta-research team introduced \texttt{CyberSecEval} to benchmark LLMs from the perspective of security \cite{bhatt2023purple}, but their prompts are in natural language and used a static analyzer to detect the vulnerabilities in the generated code.
In our work, we manually created test cases focusing on functional correctness and vulnerability detection to do the dynamic analysis. 
Other datasets and frameworks focused on specific vulnerabilities, such as regex denial-of-service attacks (ReDoS)~\cite{siddiq2024regex, siddiq2024regexeval} and hardware-specific vulnerabilities \cite{kande2024security}. 
There are also benchmarks for detecting LLM-generated code (\eg~GPTSniffer \cite{nguyen2024gptsniffer}), security vulnerability detection (\eg MSIVD \cite{yang2024securityvulnerabilitydetectionmultitask}), and improving reliability of the generated code (\eg Kouemo \etal~\cite{Ngassom2024}).
\section{Conclusion}\label{sec:conclusion}

In this study, we introduce \sandoval, a platform designed specifically for evaluating the capability of LLMs to produce secure code. This platform consists of three key elements: a unique dataset filled with security-focused Python prompts, a testing environment for the code produced, and novel metrics to assess model output.
Through our research, we utilized the \sandoval{} framework to assess 5 different LLMs. Our finding shows that GPT-4, despite being the best model for generating functional correct code, is not generating the most secure code. 






\bibliographystyle{ACM-Reference-Format}
\bibliography{references}


\begin{thebibliography}{85}


\ifx \showCODEN    \undefined \def \showCODEN     #1{\unskip}     \fi
\ifx \showDOI      \undefined \def \showDOI       #1{#1}\fi
\ifx \showISBNx    \undefined \def \showISBNx     #1{\unskip}     \fi
\ifx \showISBNxiii \undefined \def \showISBNxiii  #1{\unskip}     \fi
\ifx \showISSN     \undefined \def \showISSN      #1{\unskip}     \fi
\ifx \showLCCN     \undefined \def \showLCCN      #1{\unskip}     \fi
\ifx \shownote     \undefined \def \shownote      #1{#1}          \fi
\ifx \showarticletitle \undefined \def \showarticletitle #1{#1}   \fi
\ifx \showURL      \undefined \def \showURL       {\relax}        \fi
\providecommand\bibfield[2]{#2}
\providecommand\bibinfo[2]{#2}
\providecommand\natexlab[1]{#1}
\providecommand\showeprint[2][]{arXiv:#2}

\bibitem[Sta(2022)]%
        {StackOverflowDevSurvey}
 \bibinfo{year}{2022}\natexlab{}.
\newblock \bibinfo{title}{{Stack Overflow Developer Survey 2021}}.
\newblock
\newblock
\urldef\tempurl%
\url{https://insights.stackoverflow.com/survey/2021}
\showURL{%
\tempurl}
\newblock
\shownote{[Online; accessed 28. Aug. 2022]}.


\bibitem[cha(2023)]%
        {chatgpt}
 \bibinfo{year}{2023}\natexlab{}.
\newblock \bibinfo{title}{Chat completions}.
\newblock \bibinfo{howpublished}{Accessed Mar 25, 2023}.
\newblock
\urldef\tempurl%
\url{https://platform.openai.com/docs/guides/chat}
\showURL{%
\tempurl}


\bibitem[Sta(2024)]%
        {StarCoderIntendedUsage}
 \bibinfo{year}{2024}\natexlab{}.
\newblock \bibinfo{booktitle}{\emph{{bigcode/starcoder {$\cdot$} Hugging
  Face}}}.
\newblock
\urldef\tempurl%
\url{https://huggingface.co/bigcode/starcoder#intended-use}
\showURL{%
\tempurl}
\newblock
\shownote{[Online; accessed 10. Aug. 2024]}.


\bibitem[CWE(2024)]%
        {CWEMappingGuidance}
 \bibinfo{year}{2024}\natexlab{}.
\newblock \bibinfo{booktitle}{\emph{{CWE - CVE {$\rightarrow$} CWE Mapping
  "Root Cause Mapping" Guidance}}}.
\newblock
\urldef\tempurl%
\url{https://cwe.mitre.org/documents/cwe_usage/guidance.html}
\showURL{%
\tempurl}
\newblock
\shownote{[Online; accessed 10. Aug. 2024]}.


\bibitem[uni(2024)]%
        {unittest}
 \bibinfo{year}{2024}\natexlab{}.
\newblock \bibinfo{booktitle}{\emph{{unittest {\ifmmode---\else\textemdash\fi}
  Unit testing framework}}}.
\newblock
\urldef\tempurl%
\url{https://docs.python.org/3/library/unittest.html}
\showURL{%
\tempurl}
\newblock
\shownote{[Online; accessed 10. Aug. 2024]}.


\bibitem[Allamanis et~al\mbox{.}(2018)]%
        {allamanis2018survey}
\bibfield{author}{\bibinfo{person}{Miltiadis Allamanis},
  \bibinfo{person}{Earl~T Barr}, \bibinfo{person}{Premkumar Devanbu}, {and}
  \bibinfo{person}{Charles Sutton}.} \bibinfo{year}{2018}\natexlab{}.
\newblock \showarticletitle{A survey of machine learning for big code and
  naturalness}.
\newblock \bibinfo{journal}{\emph{ACM Computing Surveys (CSUR)}}
  \bibinfo{volume}{51}, \bibinfo{number}{4} (\bibinfo{year}{2018}),
  \bibinfo{pages}{1--37}.
\newblock


\bibitem[Athiwaratkun et~al\mbox{.}(2023)]%
        {MBXP}
\bibfield{author}{\bibinfo{person}{Ben Athiwaratkun},
  \bibinfo{person}{Sanjay~Krishna Gouda}, \bibinfo{person}{Zijian Wang},
  \bibinfo{person}{Xiaopeng Li}, \bibinfo{person}{Yuchen Tian},
  \bibinfo{person}{Ming Tan}, \bibinfo{person}{Wasi~Uddin Ahmad},
  \bibinfo{person}{Shiqi Wang}, \bibinfo{person}{Qing Sun},
  \bibinfo{person}{Mingyue Shang}, {et~al\mbox{.}}}
  \bibinfo{year}{2023}\natexlab{}.
\newblock \showarticletitle{Multi-lingual Evaluation of Code Generation
  Models}. In \bibinfo{booktitle}{\emph{The Eleventh International Conference
  on Learning Representations (ICLR)}}.
\newblock
\urldef\tempurl%
\url{https://openreview.net/forum?id=Bo7eeXm6An8}
\showURL{%
\tempurl}


\bibitem[Austin et~al\mbox{.}(2021)]%
        {MBPP}
\bibfield{author}{\bibinfo{person}{Jacob Austin}, \bibinfo{person}{Augustus
  Odena}, \bibinfo{person}{Maxwell Nye}, \bibinfo{person}{Maarten Bosma},
  \bibinfo{person}{Henryk Michalewski}, \bibinfo{person}{David Dohan},
  \bibinfo{person}{Ellen Jiang}, \bibinfo{person}{Carrie Cai},
  \bibinfo{person}{Michael Terry}, \bibinfo{person}{Quoc Le}, {and}
  \bibinfo{person}{Charles Sutton}.} \bibinfo{year}{2021}\natexlab{}.
\newblock \showarticletitle{Program synthesis with large language models}.
\newblock \bibinfo{journal}{\emph{arXiv preprint arXiv:2108.07732}}
  (\bibinfo{year}{2021}).
\newblock


\bibitem[Banerjee and Lavie(2005)]%
        {METEOR}
\bibfield{author}{\bibinfo{person}{Satanjeev Banerjee} {and}
  \bibinfo{person}{Alon Lavie}.} \bibinfo{year}{2005}\natexlab{}.
\newblock \showarticletitle{METEOR: An automatic metric for MT evaluation with
  improved correlation with human judgments}. In
  \bibinfo{booktitle}{\emph{Proceedings of the acl workshop on intrinsic and
  extrinsic evaluation measures for machine translation and/or summarization}}.
  \bibinfo{pages}{65--72}.
\newblock


\bibitem[Bhatt et~al\mbox{.}(2023)]%
        {bhatt2023purple}
\bibfield{author}{\bibinfo{person}{Manish Bhatt}, \bibinfo{person}{Sahana
  Chennabasappa}, \bibinfo{person}{Cyrus Nikolaidis}, \bibinfo{person}{Shengye
  Wan}, \bibinfo{person}{Ivan Evtimov}, \bibinfo{person}{Dominik Gabi},
  \bibinfo{person}{Daniel Song}, \bibinfo{person}{Faizan Ahmad},
  \bibinfo{person}{Cornelius Aschermann}, \bibinfo{person}{Lorenzo Fontana},
  {et~al\mbox{.}}} \bibinfo{year}{2023}\natexlab{}.
\newblock \showarticletitle{Purple llama cyberseceval: A secure coding
  benchmark for language models}.
\newblock \bibinfo{journal}{\emph{arXiv preprint arXiv:2312.04724}}
  (\bibinfo{year}{2023}).
\newblock


\bibitem[Brown et~al\mbox{.}(2020)]%
        {brown2020language}
\bibfield{author}{\bibinfo{person}{Tom Brown}, \bibinfo{person}{Benjamin Mann},
  \bibinfo{person}{Nick Ryder}, \bibinfo{person}{Melanie Subbiah},
  \bibinfo{person}{Jared~D Kaplan}, \bibinfo{person}{Prafulla Dhariwal},
  \bibinfo{person}{Arvind Neelakantan}, \bibinfo{person}{Pranav Shyam},
  \bibinfo{person}{Girish Sastry}, \bibinfo{person}{Amanda Askell},
  {et~al\mbox{.}}} \bibinfo{year}{2020}\natexlab{}.
\newblock \bibinfo{title}{Language Models are Few-Shot Learners}.
\newblock
\newblock
\showeprint[arxiv]{2005.14165}~[cs.CL]


\bibitem[Chandel et~al\mbox{.}(2022)]%
        {DSP}
\bibfield{author}{\bibinfo{person}{Shubham Chandel}, \bibinfo{person}{Colin~B
  Clement}, \bibinfo{person}{Guillermo Serrato}, {and} \bibinfo{person}{Neel
  Sundaresan}.} \bibinfo{year}{2022}\natexlab{}.
\newblock \showarticletitle{Training and evaluating a jupyter notebook data
  science assistant}.
\newblock \bibinfo{journal}{\emph{arXiv preprint arXiv:2201.12901}}
  (\bibinfo{year}{2022}).
\newblock


\bibitem[Chen et~al\mbox{.}(2021a)]%
        {HumanEval}
\bibfield{author}{\bibinfo{person}{Mark Chen}, \bibinfo{person}{Jerry Tworek},
  \bibinfo{person}{Heewoo Jun}, \bibinfo{person}{Qiming Yuan},
  \bibinfo{person}{Henrique~Ponde de Oliveira~Pinto}, {et~al\mbox{.}}}
  \bibinfo{year}{2021}\natexlab{a}.
\newblock \showarticletitle{Evaluating large language models trained on code}.
\newblock \bibinfo{journal}{\emph{arXiv preprint arXiv:2107.03374}}
  (\bibinfo{year}{2021}).
\newblock


\bibitem[Chen et~al\mbox{.}(2021b)]%
        {chen2021codex}
\bibfield{author}{\bibinfo{person}{Mark Chen}, \bibinfo{person}{Jerry Tworek},
  \bibinfo{person}{Heewoo Jun}, \bibinfo{person}{Qiming Yuan},
  \bibinfo{person}{Henrique~Ponde de Oliveira~Pinto}, {et~al\mbox{.}}}
  \bibinfo{year}{2021}\natexlab{b}.
\newblock \bibinfo{title}{Evaluating Large Language Models Trained on Code}.
\newblock
\newblock
\showeprint[arxiv]{2107.03374}~[cs.LG]


\bibitem[Corporation(2023)]%
        {cwe_328}
\bibfield{author}{\bibinfo{person}{The~MITRE Corporation}.}
  \bibinfo{year}{2023}\natexlab{}.
\newblock \bibinfo{title}{CWE-328: Use of Weak Hash}.
\newblock
\newblock
\urldef\tempurl%
\url{https://cwe.mitre.org/data/definitions/328.html}
\showURL{%
\tempurl}
\newblock
\shownote{[Online; accessed 30. May. 2023]}.


\bibitem[Devlin et~al\mbox{.}(2019)]%
        {bert2018}
\bibfield{author}{\bibinfo{person}{Jacob Devlin}, \bibinfo{person}{Ming-Wei
  Chang}, \bibinfo{person}{Kenton Lee}, {and} \bibinfo{person}{Kristina
  Toutanova}.} \bibinfo{year}{2019}\natexlab{}.
\newblock \showarticletitle{{BERT}: Pre-training of Deep Bidirectional
  Transformers for Language Understanding}. In
  \bibinfo{booktitle}{\emph{Proceedings of the 2019 Conference of the North
  {A}merican Chapter of the Association for Computational Linguistics: Human
  Language Technologies, Volume 1 (Long and Short Papers)}}.
  \bibinfo{publisher}{Association for Computational Linguistics},
  \bibinfo{address}{Minneapolis, Minnesota}, \bibinfo{pages}{4171--4186}.
\newblock
\urldef\tempurl%
\url{https://doi.org/10.18653/v1/N19-1423}
\showDOI{\tempurl}


\bibitem[Ding et~al\mbox{.}(2023)]%
        {ding2023static}
\bibfield{author}{\bibinfo{person}{Hantian Ding}, \bibinfo{person}{Varun
  Kumar}, \bibinfo{person}{Yuchen Tian}, \bibinfo{person}{Zijian Wang},
  \bibinfo{person}{Rob Kwiatkowski}, \bibinfo{person}{Xiaopeng Li},
  \bibinfo{person}{Murali~Krishna Ramanathan}, \bibinfo{person}{Baishakhi Ray},
  \bibinfo{person}{Parminder Bhatia}, {and} \bibinfo{person}{Sudipta
  Sengupta}.} \bibinfo{year}{2023}\natexlab{}.
\newblock \showarticletitle{A Static Evaluation of Code Completion by Large
  Language Models}. In \bibinfo{booktitle}{\emph{Proceedings of the 61st Annual
  Meeting of the Association for Computational Linguistics (Volume 5: Industry
  Track)}}. \bibinfo{publisher}{Association for Computational Linguistics},
  \bibinfo{address}{Toronto, Canada}, \bibinfo{pages}{347–360}.
\newblock
\urldef\tempurl%
\url{https://doi.org/10.18653/v1/2023.acl-industry.34}
\showDOI{\tempurl}


\bibitem[Feng et~al\mbox{.}(2020)]%
        {codebert}
\bibfield{author}{\bibinfo{person}{Zhangyin Feng}, \bibinfo{person}{Daya Guo},
  \bibinfo{person}{Duyu Tang}, \bibinfo{person}{Nan Duan},
  \bibinfo{person}{Xiaocheng Feng}, \bibinfo{person}{Ming Gong},
  \bibinfo{person}{Linjun Shou}, \bibinfo{person}{Bing Qin},
  \bibinfo{person}{Ting Liu}, \bibinfo{person}{Daxin Jiang}, {and}
  \bibinfo{person}{Ming Zhou}.} \bibinfo{year}{2020}\natexlab{}.
\newblock \showarticletitle{{C}ode{BERT}: A Pre-Trained Model for Programming
  and Natural Languages}. In \bibinfo{booktitle}{\emph{Findings of the
  Association for Computational Linguistics: EMNLP 2020}}.
  \bibinfo{publisher}{Association for Computational Linguistics},
  \bibinfo{address}{Online}, \bibinfo{pages}{1536--1547}.
\newblock
\urldef\tempurl%
\url{https://doi.org/10.18653/v1/2020.findings-emnlp.139}
\showDOI{\tempurl}


\bibitem[Gao et~al\mbox{.}(2020)]%
        {ThePileDataset}
\bibfield{author}{\bibinfo{person}{Leo Gao}, \bibinfo{person}{Stella Biderman},
  \bibinfo{person}{Sid Black}, \bibinfo{person}{Laurence Golding},
  \bibinfo{person}{Travis Hoppe}, \bibinfo{person}{Charles Foster},
  \bibinfo{person}{Jason Phang}, \bibinfo{person}{Horace He},
  \bibinfo{person}{Anish Thite}, \bibinfo{person}{Noa Nabeshima},
  \bibinfo{person}{Shawn Presser}, {and} \bibinfo{person}{Connor Leahy}.}
  \bibinfo{year}{2020}\natexlab{}.
\newblock \bibinfo{title}{The Pile: An 800GB Dataset of Diverse Text for
  Language Modeling}.
\newblock
\newblock
\showeprint[arxiv]{2101.00027}~[cs.CL]


\bibitem[Gao and Lyu(2022)]%
        {gao2022m2ts}
\bibfield{author}{\bibinfo{person}{Yuexiu Gao} {and} \bibinfo{person}{Chen
  Lyu}.} \bibinfo{year}{2022}\natexlab{}.
\newblock \showarticletitle{M2TS: Multi-Scale Multi-Modal Approach Based on
  Transformer for Source Code Summarization}. In
  \bibinfo{booktitle}{\emph{Proceedings of the 30th IEEE/ACM International
  Conference on Program Comprehension}} (Virtual Event)
  \emph{(\bibinfo{series}{ICPC '22})}. \bibinfo{publisher}{Association for
  Computing Machinery}, \bibinfo{address}{New York, NY, USA},
  \bibinfo{pages}{24–35}.
\newblock
\showISBNx{9781450392983}
\urldef\tempurl%
\url{https://doi.org/10.1145/3524610.3527907}
\showDOI{\tempurl}


\bibitem[Ghafari et~al\mbox{.}(2017)]%
        {ghafari2017security}
\bibfield{author}{\bibinfo{person}{Mohammad Ghafari}, \bibinfo{person}{Pascal
  Gadient}, {and} \bibinfo{person}{Oscar Nierstrasz}.}
  \bibinfo{year}{2017}\natexlab{}.
\newblock \showarticletitle{Security smells in android}. In
  \bibinfo{booktitle}{\emph{2017 IEEE 17th international working conference on
  source code analysis and manipulation (SCAM)}}. IEEE,
  \bibinfo{pages}{121--130}.
\newblock
\urldef\tempurl%
\url{https://doi.org/10.1109/SCAM.2017.24}
\showDOI{\tempurl}


\bibitem[Green(1969)]%
        {green69}
\bibfield{author}{\bibinfo{person}{Cordell Green}.}
  \bibinfo{year}{1969}\natexlab{}.
\newblock \showarticletitle{Application of Theorem Proving to Problem Solving}.
  In \bibinfo{booktitle}{\emph{Proc. of the 1st Intl. Joint Conf. on Artificial
  Intelligence}} (Washington, DC) \emph{(\bibinfo{series}{IJCAI'69})}.
  \bibinfo{publisher}{Morgan Kaufmann Publishers Inc.}, \bibinfo{address}{San
  Francisco, CA, USA}, \bibinfo{pages}{219–239}.
\newblock


\bibitem[Gulwani et~al\mbox{.}(2017)]%
        {gulwani2017program}
\bibfield{author}{\bibinfo{person}{Sumit Gulwani}, \bibinfo{person}{Oleksandr
  Polozov}, \bibinfo{person}{Rishabh Singh}, {et~al\mbox{.}}}
  \bibinfo{year}{2017}\natexlab{}.
\newblock \showarticletitle{Program synthesis}.
\newblock \bibinfo{journal}{\emph{Foundations and Trends{\textregistered} in
  Programming Languages}} \bibinfo{volume}{4}, \bibinfo{number}{1-2}
  (\bibinfo{year}{2017}), \bibinfo{pages}{1--119}.
\newblock


\bibitem[Hajipour et~al\mbox{.}(2024)]%
        {CodeLMSec}
\bibfield{author}{\bibinfo{person}{Hossein Hajipour}, \bibinfo{person}{Keno
  Hassler}, \bibinfo{person}{Thorsten Holz}, \bibinfo{person}{Lea Schönherr},
  {and} \bibinfo{person}{Mario Fritz}.} \bibinfo{year}{2024}\natexlab{}.
\newblock \showarticletitle{CodeLMSec Benchmark: Systematically Evaluating and
  Finding Security Vulnerabilities in Black-Box Code Language Models}. In
  \bibinfo{booktitle}{\emph{2024 IEEE Conference on Secure and Trustworthy
  Machine Learning (SaTML)}}. \bibinfo{pages}{684--709}.
\newblock
\urldef\tempurl%
\url{https://doi.org/10.1109/SaTML59370.2024.00040}
\showDOI{\tempurl}


\bibitem[Hajipour et~al\mbox{.}(2023)]%
        {hajipour2023systematically}
\bibfield{author}{\bibinfo{person}{Hossein Hajipour}, \bibinfo{person}{Thorsten
  Holz}, \bibinfo{person}{Lea Sch{\"o}nherr}, {and} \bibinfo{person}{Mario
  Fritz}.} \bibinfo{year}{2023}\natexlab{}.
\newblock \showarticletitle{Systematically Finding Security Vulnerabilities in
  Black-Box Code Generation Models}.
\newblock \bibinfo{journal}{\emph{arXiv preprint arXiv:2302.04012}}
  (\bibinfo{year}{2023}).
\newblock


\bibitem[Hendrycks et~al\mbox{.}(2021)]%
        {APPS}
\bibfield{author}{\bibinfo{person}{Dan Hendrycks}, \bibinfo{person}{Steven
  Basart}, \bibinfo{person}{Saurav Kadavath}, \bibinfo{person}{Mantas Mazeika},
  \bibinfo{person}{Akul Arora}, \bibinfo{person}{Ethan Guo},
  \bibinfo{person}{Collin Burns}, \bibinfo{person}{Samir Puranik},
  \bibinfo{person}{Horace He}, \bibinfo{person}{Dawn Song}, {and}
  \bibinfo{person}{Jacob Steinhardt}.} \bibinfo{year}{2021}\natexlab{}.
\newblock \showarticletitle{Measuring Coding Challenge Competence With {APPS}}.
\newblock \bibinfo{journal}{\emph{NeurIPS}} (\bibinfo{year}{2021}).
\newblock


\bibitem[Inc(2022)]%
        {BigQueryDataset}
\bibfield{author}{\bibinfo{person}{Google Inc}.}
  \bibinfo{year}{2022}\natexlab{}.
\newblock \bibinfo{title}{BigQuery public datasets}.
\newblock
\newblock
\urldef\tempurl%
\url{https://cloud.google.com/bigquery/public-data}
\showURL{%
\tempurl}


\bibitem[Inc.(2022a)]%
        {copilot}
\bibfield{author}{\bibinfo{person}{GitHub Inc.}}
  \bibinfo{year}{2022}\natexlab{a}.
\newblock \bibinfo{title}{GitHub Copilot : Your AI pair programmer}.
\newblock
\newblock
\urldef\tempurl%
\url{https://copilot.github.com}
\showURL{%
\tempurl}
\newblock
\shownote{[Online; accessed 10. Oct. 2022]}.


\bibitem[Inc.(2022b)]%
        {codeql}
\bibfield{author}{\bibinfo{person}{GitHub Inc.}}
  \bibinfo{year}{2022}\natexlab{b}.
\newblock \bibinfo{title}{Use of a broken or weak cryptographic hashing
  algorithm on sensitive data}.
\newblock
\newblock
\urldef\tempurl%
\url{https://codeql.github.com/codeql-query-help/python/py-weak-sensitive-data-hashing/}
\showURL{%
\tempurl}
\newblock
\shownote{[Online; accessed 30. Oct. 2022]}.


\bibitem[Iyer et~al\mbox{.}(2018)]%
        {iyer2018mapping}
\bibfield{author}{\bibinfo{person}{Srinivasan Iyer}, \bibinfo{person}{Ioannis
  Konstas}, \bibinfo{person}{Alvin Cheung}, {and} \bibinfo{person}{Luke
  Zettlemoyer}.} \bibinfo{year}{2018}\natexlab{}.
\newblock \showarticletitle{Mapping language to code in programmatic context}.
\newblock \bibinfo{journal}{\emph{arXiv preprint arXiv:1808.09588}}
  (\bibinfo{year}{2018}).
\newblock


\bibitem[Izadi et~al\mbox{.}(2022)]%
        {izadi2022codefill}
\bibfield{author}{\bibinfo{person}{Maliheh Izadi}, \bibinfo{person}{Roberta
  Gismondi}, {and} \bibinfo{person}{Georgios Gousios}.}
  \bibinfo{year}{2022}\natexlab{}.
\newblock \showarticletitle{CodeFill: Multi-token Code Completion by Jointly
  Learning from Structure and Naming Sequences}. In
  \bibinfo{booktitle}{\emph{44th {International} {Conference} on {Software}
  {Engineering} ({ICSE})}}.
\newblock


\bibitem[Kande et~al\mbox{.}(2024)]%
        {kande2024security}
\bibfield{author}{\bibinfo{person}{Rahul Kande}, \bibinfo{person}{Hammond
  Pearce}, \bibinfo{person}{Benjamin Tan}, \bibinfo{person}{Brendan
  Dolan-Gavitt}, \bibinfo{person}{Shailja Thakur}, \bibinfo{person}{Ramesh
  Karri}, {and} \bibinfo{person}{Jeyavijayan Rajendran}.}
  \bibinfo{year}{2024}\natexlab{}.
\newblock \showarticletitle{(Security) Assertions by Large Language Models}.
\newblock \bibinfo{journal}{\emph{IEEE Transactions on Information Forensics
  and Security}} (\bibinfo{year}{2024}).
\newblock


\bibitem[Kim et~al\mbox{.}(2021)]%
        {kim2021code}
\bibfield{author}{\bibinfo{person}{Seohyun Kim}, \bibinfo{person}{Jinman Zhao},
  \bibinfo{person}{Yuchi Tian}, {and} \bibinfo{person}{Satish Chandra}.}
  \bibinfo{year}{2021}\natexlab{}.
\newblock \showarticletitle{Code prediction by feeding trees to transformers}.
  In \bibinfo{booktitle}{\emph{2021 IEEE/ACM 43rd International Conference on
  Software Engineering (ICSE)}}. IEEE, \bibinfo{pages}{150--162}.
\newblock


\bibitem[Kocetkov et~al\mbox{.}(2022)]%
        {Kocetkov2022TheStack}
\bibfield{author}{\bibinfo{person}{Denis Kocetkov}, \bibinfo{person}{Raymond
  Li}, \bibinfo{person}{Loubna Ben~Allal}, \bibinfo{person}{Jia Li},
  \bibinfo{person}{Chenghao Mou}, \bibinfo{person}{Carlos Muñoz~Ferrandis},
  \bibinfo{person}{Yacine Jernite}, \bibinfo{person}{Margaret Mitchell},
  \bibinfo{person}{Sean Hughes}, \bibinfo{person}{Thomas Wolf},
  \bibinfo{person}{Dzmitry Bahdanau}, \bibinfo{person}{Leandro von Werra},
  {and} \bibinfo{person}{Harm de Vries}.} \bibinfo{year}{2022}\natexlab{}.
\newblock \showarticletitle{The Stack: 3 TB of permissively licensed source
  code}.
\newblock \bibinfo{journal}{\emph{Preprint}} (\bibinfo{year}{2022}).
\newblock


\bibitem[Kouemo~Ngassom et~al\mbox{.}(2024)]%
        {Ngassom2024}
\bibfield{author}{\bibinfo{person}{Sylvain Kouemo~Ngassom},
  \bibinfo{person}{Arghavan Moradi~Dakhel}, \bibinfo{person}{Florian Tambon},
  {and} \bibinfo{person}{Foutse Khomh}.} \bibinfo{year}{2024}\natexlab{}.
\newblock \showarticletitle{Chain of Targeted Verification Questions to Improve
  the Reliability of Code Generated by LLMs}. In
  \bibinfo{booktitle}{\emph{Proceedings of the 1st ACM International Conference
  on AI-Powered Software}} (Porto de Galinhas, Brazil)
  \emph{(\bibinfo{series}{AIware 2024})}. \bibinfo{publisher}{Association for
  Computing Machinery}, \bibinfo{address}{New York, NY, USA},
  \bibinfo{pages}{122–130}.
\newblock
\showISBNx{9798400706851}
\urldef\tempurl%
\url{https://doi.org/10.1145/3664646.3664772}
\showDOI{\tempurl}


\bibitem[Kulal et~al\mbox{.}(2019)]%
        {kulal2019spoc}
\bibfield{author}{\bibinfo{person}{Sumith Kulal}, \bibinfo{person}{Panupong
  Pasupat}, \bibinfo{person}{Kartik Chandra}, \bibinfo{person}{Mina Lee},
  \bibinfo{person}{Oded Padon}, \bibinfo{person}{Alex Aiken}, {and}
  \bibinfo{person}{Percy~S Liang}.} \bibinfo{year}{2019}\natexlab{}.
\newblock \showarticletitle{SPoC: Search-based Pseudocode to Code}. In
  \bibinfo{booktitle}{\emph{Advances in Neural Information Processing
  Systems}}, \bibfield{editor}{\bibinfo{person}{H.~Wallach},
  \bibinfo{person}{H.~Larochelle}, \bibinfo{person}{A.~Beygelzimer},
  \bibinfo{person}{F.~d\textquotesingle Alch\'{e}-Buc},
  \bibinfo{person}{E.~Fox}, {and} \bibinfo{person}{R.~Garnett}} (Eds.),
  Vol.~\bibinfo{volume}{32}. \bibinfo{publisher}{Curran Associates, Inc.}
\newblock


\bibitem[Lai et~al\mbox{.}(2022)]%
        {DS1000}
\bibfield{author}{\bibinfo{person}{Yuhang Lai}, \bibinfo{person}{Chengxi Li},
  \bibinfo{person}{Yiming Wang}, \bibinfo{person}{Tianyi Zhang},
  \bibinfo{person}{Ruiqi Zhong}, \bibinfo{person}{Luke Zettlemoyer},
  \bibinfo{person}{Scott Wen-tau Yih}, \bibinfo{person}{Daniel Fried},
  \bibinfo{person}{Sida Wang}, {and} \bibinfo{person}{Tao Yu}.}
  \bibinfo{year}{2022}\natexlab{}.
\newblock \showarticletitle{DS-1000: A Natural and Reliable Benchmark for Data
  Science Code Generation}.
\newblock \bibinfo{journal}{\emph{arXiv preprint arXiv:2211.11501}}
  (\bibinfo{year}{2022}).
\newblock


\bibitem[Le et~al\mbox{.}(2020)]%
        {Le_2021}
\bibfield{author}{\bibinfo{person}{Triet H.~M. Le}, \bibinfo{person}{Hao Chen},
  {and} \bibinfo{person}{Muhammad~Ali Babar}.} \bibinfo{year}{2020}\natexlab{}.
\newblock \showarticletitle{Deep Learning for Source Code Modeling and
  Generation: Models, Applications, and Challenges}.
\newblock \bibinfo{journal}{\emph{ACM Comput. Surv.}} \bibinfo{volume}{53},
  \bibinfo{number}{3}, Article \bibinfo{articleno}{62} (\bibinfo{date}{jun}
  \bibinfo{year}{2020}), \bibinfo{numpages}{38}~pages.
\newblock
\showISSN{0360-0300}
\urldef\tempurl%
\url{https://doi.org/10.1145/3383458}
\showDOI{\tempurl}


\bibitem[Li et~al\mbox{.}(2023)]%
        {StarCoder}
\bibfield{author}{\bibinfo{person}{Raymond Li}, \bibinfo{person}{Loubna~Ben
  Allal}, \bibinfo{person}{Yangtian Zi}, \bibinfo{person}{Niklas Muennighoff},
  \bibinfo{person}{Denis Kocetkov}, \bibinfo{person}{Chenghao Mou},
  \bibinfo{person}{Marc Marone}, \bibinfo{person}{Christopher Akiki},
  \bibinfo{person}{Jia Li}, \bibinfo{person}{Jenny Chim}, {et~al\mbox{.}}}
  \bibinfo{year}{2023}\natexlab{}.
\newblock \showarticletitle{{StarCoder}: may the source be with you!}
\newblock \bibinfo{journal}{\emph{arXiv preprint arXiv:2305.06161}}
  (\bibinfo{year}{2023}).
\newblock


\bibitem[Li et~al\mbox{.}(2022)]%
        {alphaCode}
\bibfield{author}{\bibinfo{person}{Yujia Li}, \bibinfo{person}{David Choi},
  \bibinfo{person}{Junyoung Chung}, \bibinfo{person}{Nate Kushman},
  \bibinfo{person}{Julian Schrittwieser}, \bibinfo{person}{R{\'{e} }mi
  Leblond}, \bibinfo{person}{Tom Eccles}, \bibinfo{person}{James Keeling},
  \bibinfo{person}{Felix Gimeno}, \bibinfo{person}{Agustin~Dal Lago},
  \bibinfo{person}{Thomas Hubert}, \bibinfo{person}{Peter Choy},
  \bibinfo{person}{Cyprien de Masson~d'Autume}, \bibinfo{person}{Igor
  Babuschkin}, \bibinfo{person}{Xinyun Chen}, \bibinfo{person}{Po-Sen Huang},
  \bibinfo{person}{Johannes Welbl}, \bibinfo{person}{Sven Gowal},
  \bibinfo{person}{Alexey Cherepanov}, \bibinfo{person}{James Molloy},
  \bibinfo{person}{Daniel~J. Mankowitz}, \bibinfo{person}{Esme~Sutherland
  Robson}, \bibinfo{person}{Pushmeet Kohli}, \bibinfo{person}{Nando de
  Freitas}, \bibinfo{person}{Koray Kavukcuoglu}, {and} \bibinfo{person}{Oriol
  Vinyals}.} \bibinfo{year}{2022}\natexlab{}.
\newblock \bibinfo{title}{Competition-Level Code Generation with AlphaCode}.
\newblock
\newblock
\urldef\tempurl%
\url{https://doi.org/10.48550/ARXIV.2203.07814}
\showDOI{\tempurl}


\bibitem[Lin(2004)]%
        {ROUGE}
\bibfield{author}{\bibinfo{person}{Chin-Yew Lin}.}
  \bibinfo{year}{2004}\natexlab{}.
\newblock \showarticletitle{Rouge: A package for automatic evaluation of
  summaries}. In \bibinfo{booktitle}{\emph{Text summarization branches out}}.
  \bibinfo{pages}{74--81}.
\newblock


\bibitem[Livshits and Lam(2005)]%
        {livshits2005finding}
\bibfield{author}{\bibinfo{person}{V~Benjamin Livshits} {and}
  \bibinfo{person}{Monica~S Lam}.} \bibinfo{year}{2005}\natexlab{}.
\newblock \showarticletitle{Finding Security Vulnerabilities in Java
  Applications with Static Analysis.}. In \bibinfo{booktitle}{\emph{USENIX
  security symposium}}, Vol.~\bibinfo{volume}{14}. \bibinfo{pages}{18--18}.
\newblock


\bibitem[Lu et~al\mbox{.}(2021)]%
        {codexglue}
\bibfield{author}{\bibinfo{person}{Shuai Lu}, \bibinfo{person}{Daya Guo},
  \bibinfo{person}{Shuo Ren}, \bibinfo{person}{Junjie Huang},
  \bibinfo{person}{Alexey Svyatkovskiy}, \bibinfo{person}{Ambrosio Blanco},
  \bibinfo{person}{Colin~B. Clement}, \bibinfo{person}{Dawn Drain},
  \bibinfo{person}{Daxin Jiang}, \bibinfo{person}{Duyu Tang},
  \bibinfo{person}{Ge Li}, \bibinfo{person}{Lidong Zhou},
  \bibinfo{person}{Linjun Shou}, \bibinfo{person}{Long Zhou},
  \bibinfo{person}{Michele Tufano}, \bibinfo{person}{Ming Gong},
  \bibinfo{person}{Ming Zhou}, \bibinfo{person}{Nan Duan},
  \bibinfo{person}{Neel Sundaresan}, \bibinfo{person}{Shao~Kun Deng},
  \bibinfo{person}{Shengyu Fu}, {and} \bibinfo{person}{Shujie Liu}.}
  \bibinfo{year}{2021}\natexlab{}.
\newblock \showarticletitle{CodeXGLUE: {A} Machine Learning Benchmark Dataset
  for Code Understanding and Generation}.
\newblock \bibinfo{journal}{\emph{CoRR}}  \bibinfo{volume}{abs/2102.04664}
  (\bibinfo{year}{2021}).
\newblock


\bibitem[Manna and Waldinger(1971)]%
        {manna2017}
\bibfield{author}{\bibinfo{person}{Zohar Manna} {and}
  \bibinfo{person}{Richard~J. Waldinger}.} \bibinfo{year}{1971}\natexlab{}.
\newblock \showarticletitle{Toward Automatic Program Synthesis}.
\newblock \bibinfo{journal}{\emph{Commun. ACM}} \bibinfo{volume}{14},
  \bibinfo{number}{3} (\bibinfo{date}{mar} \bibinfo{year}{1971}),
  \bibinfo{pages}{151–165}.
\newblock
\showISSN{0001-0782}
\urldef\tempurl%
\url{https://doi.org/10.1145/362566.362568}
\showDOI{\tempurl}


\bibitem[({MITRE})(2022)]%
        {mitre}
\bibfield{author}{\bibinfo{person}{The {MITRE}~Corporation ({MITRE})}.}
  \bibinfo{year}{2022}\natexlab{}.
\newblock \bibinfo{title}{Common Weakness Enumeration}.
\newblock
\newblock
\urldef\tempurl%
\url{https://cwe.mitre.org/}
\showURL{%
\tempurl}
\newblock
\shownote{[Online; accessed 18. Aug. 2022]}.


\bibitem[({MITRE})(2023)]%
        {top25}
\bibfield{author}{\bibinfo{person}{The {MITRE}~Corporation ({MITRE})}.}
  \bibinfo{year}{2023}\natexlab{}.
\newblock \bibinfo{title}{2023 CWE Top 25 Most Dangerous Software Weaknesses}.
\newblock
\newblock
\urldef\tempurl%
\url{https://cwe.mitre.org/data/definitions/1425.html}
\showURL{%
\tempurl}
\newblock
\shownote{[Online; accessed 18. Oct. 2023]}.


\bibitem[{Moradi Dakhel} et~al\mbox{.}(2023)]%
        {dakhel2023github}
\bibfield{author}{\bibinfo{person}{Arghavan {Moradi Dakhel}},
  \bibinfo{person}{Vahid Majdinasab}, \bibinfo{person}{Amin Nikanjam},
  \bibinfo{person}{Foutse Khomh}, \bibinfo{person}{Michel~C. Desmarais}, {and}
  \bibinfo{person}{Zhen Ming~(Jack) Jiang}.} \bibinfo{year}{2023}\natexlab{}.
\newblock \showarticletitle{GitHub Copilot AI pair programmer: Asset or
  Liability?}
\newblock \bibinfo{journal}{\emph{Journal of Systems and Software}}
  \bibinfo{volume}{203} (\bibinfo{year}{2023}), \bibinfo{pages}{111734}.
\newblock
\showISSN{0164-1212}
\urldef\tempurl%
\url{https://doi.org/10.1016/j.jss.2023.111734}
\showDOI{\tempurl}


\bibitem[Nguyen and Nadi(2022)]%
        {nguyen2022empirical}
\bibfield{author}{\bibinfo{person}{Nhan Nguyen} {and} \bibinfo{person}{Sarah
  Nadi}.} \bibinfo{year}{2022}\natexlab{}.
\newblock \showarticletitle{An empirical evaluation of GitHub copilot’s code
  suggestions}. In \bibinfo{booktitle}{\emph{Proceedings of the 19th
  International Conference on Mining Software Repositories}}
  \emph{(\bibinfo{series}{MSR ’22})}. \bibinfo{publisher}{Association for
  Computing Machinery}, \bibinfo{address}{New York, NY, USA},
  \bibinfo{pages}{1–5}.
\newblock
\showISBNx{978-1-4503-9303-4}
\urldef\tempurl%
\url{https://doi.org/10.1145/3524842.3528470}
\showDOI{\tempurl}


\bibitem[Nguyen et~al\mbox{.}(2024)]%
        {nguyen2024gptsniffer}
\bibfield{author}{\bibinfo{person}{Phuong~T Nguyen}, \bibinfo{person}{Juri
  Di~Rocco}, \bibinfo{person}{Claudio Di~Sipio}, \bibinfo{person}{Riccardo
  Rubei}, \bibinfo{person}{Davide Di~Ruscio}, {and}
  \bibinfo{person}{Massimiliano Di~Penta}.} \bibinfo{year}{2024}\natexlab{}.
\newblock \showarticletitle{GPTSniffer: A CodeBERT-based classifier to detect
  source code written by ChatGPT}.
\newblock \bibinfo{journal}{\emph{Journal of Systems and Software}}
  \bibinfo{volume}{214} (\bibinfo{year}{2024}), \bibinfo{pages}{112059}.
\newblock


\bibitem[Nijkamp et~al\mbox{.}(2023)]%
        {nijkamp2023codegen2}
\bibfield{author}{\bibinfo{person}{Erik Nijkamp}, \bibinfo{person}{Hiroaki
  Hayashi}, \bibinfo{person}{Caiming Xiong}, \bibinfo{person}{Silvio Savarese},
  {and} \bibinfo{person}{Yingbo Zhou}.} \bibinfo{year}{2023}\natexlab{}.
\newblock \showarticletitle{CodeGen2: Lessons for Training LLMs on Programming
  and Natural Languages}.
\newblock \bibinfo{journal}{\emph{ICLR}} (\bibinfo{year}{2023}).
\newblock


\bibitem[Nijkamp et~al\mbox{.}(2022)]%
        {Nijkamp2022ACP}
\bibfield{author}{\bibinfo{person}{Erik Nijkamp}, \bibinfo{person}{Bo Pang},
  \bibinfo{person}{Hiroaki Hayashi}, \bibinfo{person}{Lifu Tu},
  \bibinfo{person}{Huan Wang}, \bibinfo{person}{Yingbo Zhou},
  \bibinfo{person}{Silvio Savarese}, {and} \bibinfo{person}{Caiming Xiong}.}
  \bibinfo{year}{2022}\natexlab{}.
\newblock \showarticletitle{A Conversational Paradigm for Program Synthesis}.
\newblock \bibinfo{journal}{\emph{arXiv preprint}} (\bibinfo{year}{2022}).
\newblock


\bibitem[Odena et~al\mbox{.}(2021)]%
        {austin2021program}
\bibfield{author}{\bibinfo{person}{Augustus Odena}, \bibinfo{person}{Charles
  Sutton}, \bibinfo{person}{David~Martin Dohan}, \bibinfo{person}{Ellen Jiang},
  \bibinfo{person}{Henryk Michalewski}, \bibinfo{person}{Jacob Austin},
  \bibinfo{person}{Maarten~Paul Bosma}, \bibinfo{person}{Maxwell Nye},
  \bibinfo{person}{Michael Terry}, {and} \bibinfo{person}{Quoc~V. Le}.}
  \bibinfo{year}{2021}\natexlab{}.
\newblock \showarticletitle{Program Synthesis with Large Language Models}. In
  \bibinfo{booktitle}{\emph{n/a}}. \bibinfo{address}{n/a},
  \bibinfo{pages}{n/a}.
\newblock
\newblock
\shownote{n/a}.


\bibitem[OpenAI(2023)]%
        {openai2023gpt4}
\bibfield{author}{\bibinfo{person}{OpenAI}.} \bibinfo{year}{2023}\natexlab{}.
\newblock \bibinfo{title}{GPT-4 Technical Report}.
\newblock
\newblock
\showeprint[arxiv]{2303.08774}~[cs.CL]


\bibitem[paperswithcode(2024)]%
        {humanevalleaderboard}
\bibfield{author}{\bibinfo{person}{paperswithcode}.}
  \bibinfo{year}{2024}\natexlab{}.
\newblock \bibinfo{title}{Code Generation on HumanEval}.
\newblock
  \bibinfo{howpublished}{https://paperswithcode.com/sota/code-generation-on-humaneval}.
\newblock


\bibitem[Papineni et~al\mbox{.}(2002)]%
        {BLEU}
\bibfield{author}{\bibinfo{person}{Kishore Papineni}, \bibinfo{person}{Salim
  Roukos}, \bibinfo{person}{Todd Ward}, {and} \bibinfo{person}{Wei-Jing Zhu}.}
  \bibinfo{year}{2002}\natexlab{}.
\newblock \showarticletitle{{BLEU}: a method for automatic evaluation of
  machine translation}. In \bibinfo{booktitle}{\emph{Proceedings of the 40th
  annual meeting of the Association for Computational Linguistics}}.
  \bibinfo{pages}{311--318}.
\newblock


\bibitem[Pearce et~al\mbox{.}(2022)]%
        {pearce2021}
\bibfield{author}{\bibinfo{person}{Hammond Pearce}, \bibinfo{person}{Baleegh
  Ahmad}, \bibinfo{person}{Benjamin Tan}, \bibinfo{person}{Brendan
  Dolan-Gavitt}, {and} \bibinfo{person}{Ramesh Karri}.}
  \bibinfo{year}{2022}\natexlab{}.
\newblock \showarticletitle{Asleep at the Keyboard? Assessing the Security of
  GitHub Copilot’s Code Contributions}. In \bibinfo{booktitle}{\emph{2022
  IEEE Symposium on Security and Privacy (SP)}}. \bibinfo{pages}{754--768}.
\newblock
\urldef\tempurl%
\url{https://doi.org/10.1109/SP46214.2022.9833571}
\showDOI{\tempurl}


\bibitem[Perry et~al\mbox{.}(2022)]%
        {perry2022users}
\bibfield{author}{\bibinfo{person}{Neil Perry}, \bibinfo{person}{Megha
  Srivastava}, \bibinfo{person}{Deepak Kumar}, {and} \bibinfo{person}{Dan
  Boneh}.} \bibinfo{year}{2022}\natexlab{}.
\newblock \showarticletitle{Do Users Write More Insecure Code with AI
  Assistants?}
\newblock \bibinfo{journal}{\emph{arXiv preprint arXiv:2211.03622}}
  (\bibinfo{year}{2022}).
\newblock


\bibitem[Rahman et~al\mbox{.}(2019a)]%
        {rahman_seven_2019}
\bibfield{author}{\bibinfo{person}{Akond Rahman}, \bibinfo{person}{Chris
  Parnin}, {and} \bibinfo{person}{Laurie Williams}.}
  \bibinfo{year}{2019}\natexlab{a}.
\newblock \showarticletitle{The {Seven} {Sins}: {Security} {Smells} in
  {Infrastructure} as {Code} {Scripts}}. In \bibinfo{booktitle}{\emph{2019
  {IEEE}/{ACM} 41st {International} {Conference} on {Software} {Engineering}
  ({ICSE})}}. \bibinfo{publisher}{IEEE}, \bibinfo{address}{Montreal, QC,
  Canada}, \bibinfo{pages}{164--175}.
\newblock
\showISBNx{978-1-72810-869-8}
\urldef\tempurl%
\url{https://doi.org/10.1109/ICSE.2019.00033}
\showDOI{\tempurl}


\bibitem[Rahman et~al\mbox{.}(2019b)]%
        {rahman2019share}
\bibfield{author}{\bibinfo{person}{Md~Rayhanur Rahman}, \bibinfo{person}{Akond
  Rahman}, {and} \bibinfo{person}{Laurie Williams}.}
  \bibinfo{year}{2019}\natexlab{b}.
\newblock \showarticletitle{Share, But be Aware: Security Smells in Python
  Gists}. In \bibinfo{booktitle}{\emph{2019 IEEE International Conference on
  Software Maintenance and Evolution (ICSME)}}. \bibinfo{pages}{536--540}.
\newblock
\urldef\tempurl%
\url{https://doi.org/10.1109/ICSME.2019.00087}
\showDOI{\tempurl}


\bibitem[Ren et~al\mbox{.}(2020)]%
        {CodeBLEU}
\bibfield{author}{\bibinfo{person}{Shuo Ren}, \bibinfo{person}{Daya Guo},
  \bibinfo{person}{Shuai Lu}, \bibinfo{person}{Long Zhou},
  \bibinfo{person}{Shujie Liu}, \bibinfo{person}{Duyu Tang},
  \bibinfo{person}{Neel Sundaresan}, \bibinfo{person}{Ming Zhou},
  \bibinfo{person}{Ambrosio Blanco}, {and} \bibinfo{person}{Shuai Ma}.}
  \bibinfo{year}{2020}\natexlab{}.
\newblock \showarticletitle{{CodeBLEU}: a method for automatic evaluation of
  code synthesis}.
\newblock \bibinfo{journal}{\emph{arXiv preprint arXiv:2009.10297}}
  (\bibinfo{year}{2020}).
\newblock


\bibitem[Roziere et~al\mbox{.}(2024)]%
        {CodeLLAMA}
\bibfield{author}{\bibinfo{person}{Baptiste Roziere}, \bibinfo{person}{Jonas
  Gehring}, \bibinfo{person}{Fabian Gloeckle}, \bibinfo{person}{Sten Sootla},
  \bibinfo{person}{Itai Gat}, \bibinfo{person}{Xiaoqing~Ellen Tan},
  \bibinfo{person}{Yossi Adi}, \bibinfo{person}{Jingyu Liu},
  \bibinfo{person}{Tal Remez}, \bibinfo{person}{J{\'e}r{\'e}my Rapin},
  {et~al\mbox{.}}} \bibinfo{year}{2024}\natexlab{}.
\newblock \bibinfo{title}{Code Llama: Open Foundation Models for Code}.
\newblock
\newblock
\showeprint[arxiv]{2308.12950}~[cs.CL]


\bibitem[S.A(2022)]%
        {sonar}
\bibfield{author}{\bibinfo{person}{{SonarSource} S.A}.}
  \bibinfo{year}{2022}\natexlab{}.
\newblock \bibinfo{title}{{SonarSource} static code analysis}.
\newblock \bibinfo{howpublished}{https://rules.sonarsource.com}.
\newblock


\bibitem[Sandoval et~al\mbox{.}(2022)]%
        {sandoval2022security}
\bibfield{author}{\bibinfo{person}{Gustavo Sandoval}, \bibinfo{person}{Hammond
  Pearce}, \bibinfo{person}{Teo Nys}, \bibinfo{person}{Ramesh Karri},
  \bibinfo{person}{Brendan Dolan-Gavitt}, {and} \bibinfo{person}{Siddharth
  Garg}.} \bibinfo{year}{2022}\natexlab{}.
\newblock \showarticletitle{Security Implications of Large Language Model Code
  Assistants: A User Study}.
\newblock \bibinfo{journal}{\emph{arXiv preprint arXiv:2208.09727}}
  (\bibinfo{year}{2022}).
\newblock


\bibitem[Schwartz et~al\mbox{.}(2010)]%
        {schwartz2010all}
\bibfield{author}{\bibinfo{person}{Edward~J Schwartz},
  \bibinfo{person}{Thanassis Avgerinos}, {and} \bibinfo{person}{David
  Brumley}.} \bibinfo{year}{2010}\natexlab{}.
\newblock \showarticletitle{All you ever wanted to know about dynamic taint
  analysis and forward symbolic execution (but might have been afraid to ask)}.
  In \bibinfo{booktitle}{\emph{2010 IEEE symposium on Security and privacy}}.
  IEEE, \bibinfo{pages}{317--331}.
\newblock


\bibitem[Shani(2023)]%
        {shani2023survey}
\bibfield{author}{\bibinfo{person}{Inbal Shani}.}
  \bibinfo{year}{2023}\natexlab{}.
\newblock \showarticletitle{{Survey reveals AI{'}s impact on the developer
  experience {$\vert$} The GitHub Blog}}.
\newblock \bibinfo{journal}{\emph{GitHub Blog}} (\bibinfo{date}{June}
  \bibinfo{year}{2023}).
\newblock
\urldef\tempurl%
\url{https://github.blog/2023-06-13-survey-reveals-ais-impact-on-the-developer-experience/#methodology}
\showURL{%
\tempurl}


\bibitem[Siddiq et~al\mbox{.}(2023)]%
        {siddiq2023franc}
\bibfield{author}{\bibinfo{person}{Mohammed~Latif Siddiq},
  \bibinfo{person}{Beatrice Casey}, {and} \bibinfo{person}{Joanna Santos}.}
  \bibinfo{year}{2023}\natexlab{}.
\newblock \showarticletitle{A Lightweight Framework for High-Quality Code
  Generation}.
\newblock \bibinfo{journal}{\emph{arXiv preprint arXiv:2307.08220}}
  (\bibinfo{year}{2023}).
\newblock


\bibitem[Siddiq et~al\mbox{.}(2022)]%
        {siddiq2022empirical}
\bibfield{author}{\bibinfo{person}{Mohammed~Latif Siddiq},
  \bibinfo{person}{Shafayat~Hossain Majumder}, \bibinfo{person}{Maisha~Rahman
  Mim}, \bibinfo{person}{Sourov Jajodia}, {and} \bibinfo{person}{Joanna~C.S.
  Santos}.} \bibinfo{year}{2022}\natexlab{}.
\newblock \showarticletitle{An Empirical Study of Code Smells in
  Transformer-based Code Generation Techniques}. In
  \bibinfo{booktitle}{\emph{2022 IEEE 22nd International Working Conference on
  Source Code Analysis and Manipulation (SCAM)}}.
\newblock


\bibitem[Siddiq et~al\mbox{.}(2024a)]%
        {siddiq2024devgpt}
\bibfield{author}{\bibinfo{person}{Mohammed~Latif Siddiq},
  \bibinfo{person}{Lindsay Roney}, \bibinfo{person}{Jiahao Zhang}, {and}
  \bibinfo{person}{Joanna C.~S. Santos}.} \bibinfo{year}{2024}\natexlab{a}.
\newblock \showarticletitle{Quality Assessment of ChatGPT Generated Code and
  their Use by Developers}. In \bibinfo{booktitle}{\emph{Proceedings of the
  21st International Conference on Mining Software Repositories, Mining
  Challenge Track (MSR 2024)}}.
\newblock


\bibitem[Siddiq and Santos(2022)]%
        {siddiq2022seceval}
\bibfield{author}{\bibinfo{person}{Mohammed~Latif Siddiq} {and}
  \bibinfo{person}{Joanna C.~S. Santos}.} \bibinfo{year}{2022}\natexlab{}.
\newblock \showarticletitle{SecurityEval Dataset: Mining Vulnerability Examples
  to Evaluate Machine Learning-Based Code Generation Techniques}. In
  \bibinfo{booktitle}{\emph{Proceedings of the 1st International Workshop on
  Mining Software Repositories Applications for Privacy and Security
  (MSR4P\&S22)}}.
\newblock
\urldef\tempurl%
\url{https://doi.org/10.1145/3549035.3561184}
\showDOI{\tempurl}


\bibitem[Siddiq et~al\mbox{.}(2024b)]%
        {siddiq2023exploring}
\bibfield{author}{\bibinfo{person}{Mohammed~Latif Siddiq},
  \bibinfo{person}{Joanna C.~S. Santos}, \bibinfo{person}{Ridwanul~Hasan
  Tanvir}, \bibinfo{person}{Noshin Ulfat}, \bibinfo{person}{Fahmid~Al Rifat},
  {and} \bibinfo{person}{Vinicius~Carvalho Lopes}.}
  \bibinfo{year}{2024}\natexlab{b}.
\newblock \showarticletitle{Using Large Language Models to Generate JUnit
  Tests: An Empirical Study}. In \bibinfo{booktitle}{\emph{28th International
  Conference on Evaluation and Assessment in Software Engineering (EASE
  2024)}}.
\newblock


\bibitem[Siddiq et~al\mbox{.}(2024d)]%
        {siddiq2024regexeval}
\bibfield{author}{\bibinfo{person}{Mohammed~Latif Siddiq},
  \bibinfo{person}{Jiahao Zhang}, \bibinfo{person}{Lindsay Roney}, {and}
  \bibinfo{person}{Joanna C.~S. Santos}.} \bibinfo{year}{2024}\natexlab{d}.
\newblock \showarticletitle{Re(gEx|DoS)Eval: Evaluating Generated Regular
  Expressions and their Proneness to DoS Attacks}. In
  \bibinfo{booktitle}{\emph{Proceedings of the 46th International Conference on
  Software Engineering, NIER Track (ICSE-NIER '24)}}.
\newblock


\bibitem[Siddiq et~al\mbox{.}(2024c)]%
        {siddiq2024regex}
\bibfield{author}{\bibinfo{person}{Mohammed~Latif Siddiq},
  \bibinfo{person}{Jiahao Zhang}, {and} \bibinfo{person}{Joanna C.~S. Santos}.}
  \bibinfo{year}{2024}\natexlab{c}.
\newblock \showarticletitle{Understanding Regular Expression Denial of Service
  (ReDoS): Insights from LLM-Generated Regexes and Developer Forums}. In
  \bibinfo{booktitle}{\emph{32nd IEEE/ACM International Conference on Program
  Comprehension (ICPC 2024)}}.
\newblock
\urldef\tempurl%
\url{https://doi.org/10.1145/3643916.3644424}
\showDOI{\tempurl}


\bibitem[Sobania et~al\mbox{.}(2022)]%
        {sobania2022choose}
\bibfield{author}{\bibinfo{person}{Dominik Sobania}, \bibinfo{person}{Martin
  Briesch}, {and} \bibinfo{person}{Franz Rothlauf}.}
  \bibinfo{year}{2022}\natexlab{}.
\newblock \showarticletitle{Choose your programming copilot: a comparison of
  the program synthesis performance of github copilot and genetic programming}.
  In \bibinfo{booktitle}{\emph{Proceedings of the Genetic and Evolutionary
  Computation Conference}} \emph{(\bibinfo{series}{GECCO ’22})}.
  \bibinfo{publisher}{Association for Computing Machinery},
  \bibinfo{address}{New York, NY, USA}, \bibinfo{pages}{1019–1027}.
\newblock
\showISBNx{978-1-4503-9237-2}
\urldef\tempurl%
\url{https://doi.org/10.1145/3512290.3528700}
\showDOI{\tempurl}


\bibitem[Svyatkovskiy et~al\mbox{.}(2021)]%
        {svyatkovskiy2021fast}
\bibfield{author}{\bibinfo{person}{Alexey Svyatkovskiy},
  \bibinfo{person}{Sebastian Lee}, \bibinfo{person}{Anna Hadjitofi},
  \bibinfo{person}{Maik Riechert}, \bibinfo{person}{Juliana~Vicente Franco},
  {and} \bibinfo{person}{Miltiadis Allamanis}.}
  \bibinfo{year}{2021}\natexlab{}.
\newblock \showarticletitle{Fast and memory-efficient neural code completion}.
  In \bibinfo{booktitle}{\emph{2021 IEEE/ACM 18th International Conference on
  Mining Software Repositories (MSR)}}. IEEE, \bibinfo{pages}{329--340}.
\newblock


\bibitem[{The MITRE Corporation}(2024)]%
        {cwe_918}
\bibfield{author}{\bibinfo{person}{{The MITRE Corporation}}.}
  \bibinfo{year}{2024}\natexlab{}.
\newblock \bibinfo{title}{CWE-918: Server-Side Request Forgery (SSRF) (4.15)}.
\newblock
  \bibinfo{howpublished}{https://cwe.mitre.org/data/definitions/918.html}.
\newblock
\newblock
\shownote{[Online; accessed 10. Aug. 2024]}.


\bibitem[Tony et~al\mbox{.}(2023)]%
        {llmseceval}
\bibfield{author}{\bibinfo{person}{C. Tony}, \bibinfo{person}{M. Mutas},
  \bibinfo{person}{N. Ferreyra}, {and} \bibinfo{person}{R. Scandariato}.}
  \bibinfo{year}{2023}\natexlab{}.
\newblock \showarticletitle{LLMSecEval: A Dataset of Natural Language Prompts
  for Security Evaluations}. In \bibinfo{booktitle}{\emph{2023 IEEE/ACM 20th
  International Conference on Mining Software Repositories (MSR)}}.
  \bibinfo{publisher}{IEEE Computer Society}, \bibinfo{address}{Los Alamitos,
  CA, USA}, \bibinfo{pages}{588--592}.
\newblock
\urldef\tempurl%
\url{https://doi.org/10.1109/MSR59073.2023.00084}
\showDOI{\tempurl}


\bibitem[Vaswani et~al\mbox{.}(2017)]%
        {attention2017}
\bibfield{author}{\bibinfo{person}{Ashish Vaswani}, \bibinfo{person}{Noam
  Shazeer}, \bibinfo{person}{Niki Parmar}, \bibinfo{person}{Jakob Uszkoreit},
  \bibinfo{person}{Llion Jones}, \bibinfo{person}{Aidan~N Gomez},
  \bibinfo{person}{\L~ukasz Kaiser}, {and} \bibinfo{person}{Illia Polosukhin}.}
  \bibinfo{year}{2017}\natexlab{}.
\newblock \showarticletitle{Attention is All you Need}. In
  \bibinfo{booktitle}{\emph{Advances in Neural Information Processing
  Systems}}, \bibfield{editor}{\bibinfo{person}{I.~Guyon},
  \bibinfo{person}{U.~Von Luxburg}, \bibinfo{person}{S.~Bengio},
  \bibinfo{person}{H.~Wallach}, \bibinfo{person}{R.~Fergus},
  \bibinfo{person}{S.~Vishwanathan}, {and} \bibinfo{person}{R.~Garnett}}
  (Eds.), Vol.~\bibinfo{volume}{30}. \bibinfo{publisher}{Curran Associates,
  Inc.}
\newblock


\bibitem[Wang et~al\mbox{.}(2021)]%
        {codet5}
\bibfield{author}{\bibinfo{person}{Yue Wang}, \bibinfo{person}{Weishi Wang},
  \bibinfo{person}{Shafiq Joty}, {and} \bibinfo{person}{Steven~C.H. Hoi}.}
  \bibinfo{year}{2021}\natexlab{}.
\newblock \showarticletitle{{C}ode{T}5: Identifier-aware Unified Pre-trained
  Encoder-Decoder Models for Code Understanding and Generation}. In
  \bibinfo{booktitle}{\emph{Proceedings of the 2021 Conference on Empirical
  Methods in Natural Language Processing}}. \bibinfo{publisher}{Association for
  Computational Linguistics}, \bibinfo{address}{Online and Punta Cana,
  Dominican Republic}, \bibinfo{pages}{8696--8708}.
\newblock
\urldef\tempurl%
\url{https://doi.org/10.18653/v1/2021.emnlp-main.685}
\showDOI{\tempurl}


\bibitem[Yamaguchi et~al\mbox{.}(2015)]%
        {yamaguchi2015automatic}
\bibfield{author}{\bibinfo{person}{Fabian Yamaguchi}, \bibinfo{person}{Alwin
  Maier}, \bibinfo{person}{Hugo Gascon}, {and} \bibinfo{person}{Konrad Rieck}.}
  \bibinfo{year}{2015}\natexlab{}.
\newblock \showarticletitle{Automatic inference of search patterns for
  taint-style vulnerabilities}. In \bibinfo{booktitle}{\emph{2015 IEEE
  Symposium on Security and Privacy}}. IEEE, \bibinfo{pages}{797--812}.
\newblock


\bibitem[Yang et~al\mbox{.}(2024)]%
        {yang2024securityvulnerabilitydetectionmultitask}
\bibfield{author}{\bibinfo{person}{Aidan Z.~H. Yang}, \bibinfo{person}{Haoye
  Tian}, \bibinfo{person}{He Ye}, \bibinfo{person}{Ruben Martins}, {and}
  \bibinfo{person}{Claire~Le Goues}.} \bibinfo{year}{2024}\natexlab{}.
\newblock \bibinfo{title}{Security Vulnerability Detection with Multitask
  Self-Instructed Fine-Tuning of Large Language Models}.
\newblock
\newblock
\showeprint[arxiv]{2406.05892}~[cs.CR]
\urldef\tempurl%
\url{https://arxiv.org/abs/2406.05892}
\showURL{%
\tempurl}


\bibitem[Yu et~al\mbox{.}(2023a)]%
        {CoderEval}
\bibfield{author}{\bibinfo{person}{Hao Yu}, \bibinfo{person}{Bo Shen},
  \bibinfo{person}{Dezhi Ran}, \bibinfo{person}{Jiaxin Zhang},
  \bibinfo{person}{Qi Zhang}, \bibinfo{person}{Yuchi Ma},
  \bibinfo{person}{Guangtai Liang}, \bibinfo{person}{Ying Li},
  \bibinfo{person}{Tao Xie}, {and} \bibinfo{person}{Qianxiang Wang}.}
  \bibinfo{year}{2023}\natexlab{a}.
\newblock \showarticletitle{CoderEval: A Benchmark of Pragmatic Code Generation
  with Generative Pre-trained Models}.
\newblock \bibinfo{journal}{\emph{arXiv preprint arXiv:2302.00288}}
  (\bibinfo{year}{2023}).
\newblock


\bibitem[Yu et~al\mbox{.}(2023b)]%
        {yu2023codereval}
\bibfield{author}{\bibinfo{person}{Hao Yu}, \bibinfo{person}{Bo Shen},
  \bibinfo{person}{Dezhi Ran}, \bibinfo{person}{Jiaxin Zhang},
  \bibinfo{person}{Qi Zhang}, \bibinfo{person}{Yuchi Ma},
  \bibinfo{person}{Guangtai Liang}, \bibinfo{person}{Ying Li},
  \bibinfo{person}{Tao Xie}, {and} \bibinfo{person}{Qianxiang Wang}.}
  \bibinfo{year}{2023}\natexlab{b}.
\newblock \bibinfo{title}{CoderEval: A Benchmark of Pragmatic Code Generation
  with Generative Pre-trained Models}.
\newblock
\newblock
\showeprint[arxiv]{2302.00288}~[cs.SE]


\bibitem[Zan et~al\mbox{.}(2023)]%
        {zan2023NL2Code}
\bibfield{author}{\bibinfo{person}{Daoguang Zan}, \bibinfo{person}{Bei Chen},
  \bibinfo{person}{Fengji Zhang}, \bibinfo{person}{Dianjie Lu},
  \bibinfo{person}{Bingchao Wu}, \bibinfo{person}{Bei Guan},
  \bibinfo{person}{Yongji Wang}, {and} \bibinfo{person}{Jian-Guang Lou}.}
  \bibinfo{year}{2023}\natexlab{}.
\newblock \showarticletitle{When Neural Model Meets {NL2Code}: A Survey}. In
  \bibinfo{booktitle}{\emph{Proceedings of the 61st Annual Meeting of the
  Association for Computational Linguistics}}.
\newblock


\bibitem[Zheng et~al\mbox{.}(2023)]%
        {zheng2023codegeex}
\bibfield{author}{\bibinfo{person}{Qinkai Zheng}, \bibinfo{person}{Xiao Xia},
  \bibinfo{person}{Xu Zou}, \bibinfo{person}{Yuxiao Dong},
  \bibinfo{person}{Shan Wang}, \bibinfo{person}{Yufei Xue},
  \bibinfo{person}{Zihan Wang}, \bibinfo{person}{Lei Shen},
  \bibinfo{person}{Andi Wang}, \bibinfo{person}{Yang Li}, \bibinfo{person}{Teng
  Su}, \bibinfo{person}{Zhilin Yang}, {and} \bibinfo{person}{Jie Tang}.}
  \bibinfo{year}{2023}\natexlab{}.
\newblock \bibinfo{title}{CodeGeeX: A Pre-Trained Model for Code Generation
  with Multilingual Evaluations on HumanEval-X}.
\newblock
\newblock
\showeprint[arxiv]{2303.17568}~[cs.LG]


\bibitem[Ziegler et~al\mbox{.}(2022)]%
        {albert22}
\bibfield{author}{\bibinfo{person}{Albert Ziegler}, \bibinfo{person}{Eirini
  Kalliamvakou}, \bibinfo{person}{X.~Alice Li}, \bibinfo{person}{Andrew Rice},
  \bibinfo{person}{Devon Rifkin}, \bibinfo{person}{Shawn Simister},
  \bibinfo{person}{Ganesh Sittampalam}, {and} \bibinfo{person}{Edward
  Aftandilian}.} \bibinfo{year}{2022}\natexlab{}.
\newblock \showarticletitle{Productivity Assessment of Neural Code Completion}.
  In \bibinfo{booktitle}{\emph{Proceedings of the 6th ACM SIGPLAN Int’l
  Symposium on Machine Programming}} (San Diego, CA, USA)
  \emph{(\bibinfo{series}{MAPS 2022})}. \bibinfo{publisher}{ACM},
  \bibinfo{address}{New York, NY, USA}, \bibinfo{pages}{21–29}.
\newblock
\showISBNx{9781450392730}
\urldef\tempurl%
\url{https://doi.org/10.1145/3520312.3534864}
\showDOI{\tempurl}


\end{thebibliography}

\end{document}